\documentclass{aa}  

\usepackage{graphicx}
\usepackage{txfonts}
\usepackage{hyperref}
\usepackage{times}
\usepackage{amsmath}
\usepackage{amssymb}
\usepackage{natbib}
\usepackage{babel}
\usepackage{xspace}
\usepackage{array}
\usepackage{multirow}
\bibpunct{(}{)}{;}{a}{}{,}

\newcommand{\mrm}{\mathrm}
\newcommand {\be} {\begin {equation}}
\newcommand {\ee} {\end {equation}}

\newcommand{\msun}{\mrm{M}_\odot}

\newcommand{\ben}{\begin{eqnarray}}
\newcommand{\een}{\end{eqnarray}}

\newcommand {\vgap} {\noalign{\vspace{0.5mm}}}

\begin{document} 

   \title{Global survey of star clusters in the Milky Way VIII. Cluster formation and evolution}


\author{J.H.~Klos\inst{1} \and
        A.~Just\inst{1} \and 
        E.V.~Polyachenko\inst{1,2,3} \and
        P.~Berczik\inst{2,4,5,6} \and
        M.~Ishchenko\inst{2,5,6}
        }
\institute{
Zentrum f\"ur Astronomie der Universit\"at
Heidelberg, Astronomisches Rechen-Institut, M\"{o}nchhofstra\ss{}e 12-14, 69120 Heidelberg, Germany
\and
Nicolaus Copernicus Astronomical Centre Polish Academy of Sciences, ul. Bartycka 18, 00-716 Warsaw, Poland
\and
SnT SEDAN, University of Luxembourg, 29 boulevard JF Kennedy, 1855 Luxembourg, Luxembourg
\and
Konkoly Observatory, HUN-REN Research Centre for Astronomy and Earth Sciences, Konkoly Thege Mikl\'os \'ut 15-17, 1121 Budapest, Hungary
\and
Fesenkov Astrophysical Institute, 23 Observatory str., 050020 Almaty, Kazakhstan
\and
Main Astronomical Observatory, National Academy of Sciences of Ukraine, 27 Akademika Zabolotnoho St., 03680, Kyiv, Ukraine
}

\date{Received ... / Accepted }

\abstract 
{We consider tidal masses and ages of Milky Way open clusters, as well as a simple model of their distribution. This model is presented as part of the Milky Way Star Cluster (MWSC) survey.} 
{Our aim is to investigate the space of model parameters and the correspondence between modelled and observed two-dimensional (2D) cluster age-mass distributions.}
{The model for cluster evolution is comprised of a two-section cluster initial mass function, constant cluster formation rate, and a mass loss function. This mass loss function represents a supervirial phase after sudden expulsion of the remaining gas, cluster mass loss due to stellar evolution and gradual cluster dissolution driven by internal dynamics and the Galactic tidal field. We constructed different estimators of model fitness based on $\chi^2$-statistics, the Kullback-Leibler divergence (KLD) and a maximum-likelihood approach, taking into account the uncertainty of our observed cluster parameters. Using these estimators and Markov chain Monte Carlo (MCMC) sampling, we obtained best-fit values and posterior distributions for a selection of model parameters.}
{The KLD returned a superior model compared to the other statistics, because it also reproduced the low-density regions of the observed cluster age-mass distribution. The cluster initial mass function is well constrained and we find a clear signature of an enhanced cluster mass loss in the first 50 Myr. Deviations from a constant cluster formation rate could not be determined due to its strong degeneracy with the shape of the cluster mass loss function. In the KLD best model,
clusters lose
{72\%} of their initial mass {in the violent relaxation phase,}
after which cluster mass loss slows down, allowing for a relatively low rate of cluster formation of ${0.088\msun \mrm{kpc^{-2} Gyr^{-1}}}$. The observed upper limit of cluster ages at approx.~5~Gyr is reflected in the model by a very shallow lifetime-mass relation for clusters with initial masses above 1000$\msun$.  {The application of the model to an independent cluster sample based on Gaia DR3 data yielded similar results except for a systematic shift in typical age and higher number densities.}}
{We conclude that the observed cluster age-mass distribution is compatible with a constant cluster formation rate. Strong correlations between model parameters reflect a sensitive dependence of the cluster age-mass distribution not just on the formation rate and initial mass function, but the details of cluster mass loss and dissolution in particular.
The enhanced number of young massive clusters observed requires an early violent relaxation phase of strong mass loss. The cluster age limit cannot be fully explained by an initial mass cutoff.}

\keywords{
Galaxy: evolution --
Galaxy: open clusters and associations: general --
Galaxy: stellar content --
Galaxies: fundamental parameters --
Galaxies: star clusters}
    
\titlerunning{MWSC VIII. Cluster formation and evolution}

\maketitle
%

\section{Introduction}\label{sec:intro}

Two of the most fundamental parameters of a star cluster are its age and mass. 
They become of particular interest in the study of star cluster populations. The age structure of a cluster system bears the imprint of its history of cluster formation and dissolution, which links to the star formation history of the host galaxy. As the cluster mass changes across a cluster's lifetime due to stellar evolution and the loss of member stars, the cluster mass function (CMF) depends on not only the conditions under which clusters form, but also on the cluster ages and lifetimes. In order to reconstruct the cluster formation rate (CFR) and the cluster initial mass function (CIMF) it would be best to use the full 2D distribution of clusters in the age-mass plane. However, literature is scarce on quantitative analysis or modelling of the full 2D {cluster age-mass function (CAMF)}. Main reasons are the sparse population of the age-mass plane and the inhomogeneity of the datasets.

If only the total cluster age function (CAF) or CMFs in some age ranges are available, a significant amount of information is hidden, which results in a large amount of ambiguity.
Typically, the CAF is used to fit and evaluate cluster evolution models quantitatively \citep[e.g.][]{lamea,mwscage}.
In another approach, \citet{2008A&A...482..165G} linked the maximum cluster mass observed for different ages to the CIMF and the rate of mass-independent cluster disruption. \citet{2010ApJ...712..604E} discussed how observed regularities in the number of clusters within a fixed mass range for different age bins reflect features of cluster evolution. They used random sampling of different cluster evolution models to qualitatively reproduce these regularities.
For extragalactic cluster systems, detection limits constrain the data base additionally to the high-mass end \citep[see e.g.][for the LMC and M83, respectively]{larss09,foue12}.

Recently, \citet{mwscmass} derived tidal masses for open clusters of the MWSC catalogue \citep[see][]{khea12,mwscat,mwscnew,mwscnew2} and constructed CMFs for different cluster age ranges for a completeness-corrected sample of Galactic open clusters. They described the CMFs in terms of a low-mass and a high-mass power-law slope, and find that while the high-mass slope remains close to one for all ages, the low-mass slope is initially shallow for clusters younger than 20~Myr. For older clusters, the low-mass slope becomes approximately $-0.7$ and changes little after the first 20~Myr. As the cluster ages increase, {the breakpoint of} this low-mass slope is shifted to higher masses. 
They present a simple analytical model for cluster formation and evolution able to reproduce the general shape and several features of these CMFs, in particular the high-mass and low-mass slopes for ages above 20~Myr, as well as the shift of the CMFs towards higher masses with increasing age.

It is our aim to investigate this model now in the context of the full 2D CAMF, which describes the Milky Way cluster surface density as a function in the cluster age-mass plane. We construct the empirical CAMF and provide a quantitative analysis of the model's correspondence to observations. Further, we perform a study of the model parameter space in order to find better estimates of the model parameters and related quantities of interest in order to improve the quality of the model as a description of the observed cluster population.

In Sect.~\ref{sec:data}, we use the MWSC cluster sample \citep{mwscat} to construct the observed CAMF for the solar neighbourhood after correcting for completeness using the approach of \citet{mwscage}. Two different views on the CAMF are presented taking into account uncertainties in the observed ages and masses.
We proceed in Sect.~\ref{sec:model} to describe and motivate the analytical CAMF model first introduced in \citet{mwscmass}, which is composed of functions governing the rate of cluster formation, the cluster initial mass distribution and the evolution of the cluster bound mass.
In Sect.~\ref{sec:methods}, we describe the methods used to artificially introduce age-mass uncertainties to the model and to fit the model parameters using a MCMC approach. The results of this approach are discussed in Sect.~\ref{sec:results}, where we compare different fit statistics used and analyse our best-fitting model.
{In Sect.~\ref{sec:gaia_data}, we compare our model to the cluster catalogue based on Gaia data of \citet{2023A&A...673A.114H}. }
Finally, Sect.~\ref{sec:conc} contains a summary of our results and conclusions.

\section{Data}\label{sec:data}

{In contrast to cluster ages, data on cluster masses for Milky Way clusters is not easy to compute and therefore not very abundant. One can mention the earlier data of \cite{lala03} based on a collection of data of 76 embedded clusters; of \cite{lamea} and of \cite{clumart07} for 520 clusters from early release of the Catalog of Open Cluster Data \citep[COCD,][]{clucat}, and of \cite{clumart} for 650 clusters from the final release of COCD. Later \cite{mwscat} based on MWSC data determined tidal radii and masses \citep{mwscmass} for 3061 open clusters. The recent releases of the Gaia survey \citep{gaiamission_16} allow for new determinations of cluster masses based on the DR2 and (E)DR3 releases: \cite{meingast21} for ten nearby clusters, \cite{cordoniea23} for 78 clusters, \cite{almeidaea23} for 773 objects, and by \cite{hunt&reffert24a} for 6956 clusters previously identified by them.

The most commonly used approach to mass determination is the derivation of the so-called counted mass via the summation of the masses of individual apparent cluster members \citep[as in][]{lala03,lamea,meingast21,cordoniea23,almeidaea23,hunt&reffert24a}. Usually, direct summation of observed individual member masses is accompanied by extrapolation of the constructed member mass function (or its template) to the arbitrarily chosen lower mass limit. The arbitrary use of the template parameters (the slope and the limit) could lead to a serious distance-related bias of the calculated mass.
Further, the measurement of cluster member masses typically depends on the cluster age, especially for the most luminous, fastest-evolving stars.

The measurement of tidal radii and computation of the corresponding bound masses using a model of the Galactic gravitational potential \citep{clumart,mwscat} is an alternative which is free from the above bias, and is probably the only direct measure of cluster mass available in practice. 
This has only become possible with the publication of all-sky surveys such as Hipparcos/Tycho \citep{ESA97}, the Two Micron All Sky Survey \citep[2MASS,][]{cat2mass}, the PPMXL Catalog of Positions and Proper Motions on the ICRS \citep[PPMXL,][]{ppmxl}, and Gaia \citep[e.g.][]{gaiadr2_18,gaiaedr3_21,gaiadr3_23}. Their use opens the possibility to determine extensive lists of cluster tidal radii and masses, which can be used as an independent source of data for the CMF.
It should be noted, however, that the determination of tidal masses is not without its own difficulties;
such as difficulties with fixing the cluster outer boundary in dense or variable extinction environments. However, the independence from the age determination and weaker assumptions about the behaviour of the unseen fraction of the cluster population compensate for this drawback.

Once the cluster masses are fixed, one can construct the CMF, or its 2D analogue, the age-mass distribution. The main requirement is full statistical representativeness of the cluster sample over the entire mass range. {This means that the sample should provide a sufficient number of objects even at the extremes of the mass and age scales. {In practice, a sample of several hundred clusters at the least is needed to resolve interesting features in the age-mass plane.} Further, the application of a proper data completeness model to the raw distributions is necessary to account for varying completeness at different masses and ages.} These conditions cut off all the poorly populated samples mentioned above. The \citet{hunt&reffert24a} sample represents the most populated list of cluster masses and ages ever published. {However, the cluster age and mass distributions derived from these data are based only on a provisional, mass-limited completeness model.}
The magnitude-limited nature of the Gaia survey suggests a completeness model based on cluster luminosities, for which a completeness distance determined solely by cluster masses is only a very limited substitute. However, cluster luminosities are neither constructed nor discussed by \citet{hunt&reffert24a}{, and no follow-up studies of the catalogue's completeness have been published at the time of writing}.

Though comparison with Gaia-based cluster catalogues \citep[such as in][]{2023A&A...673A.114H} give strong hints that the MWSC is not complete at any distance, full completeness is not essential for analysis of the CAMF as long as a representative cluster sample can be obtained.
For validation of the completeness of the MWSC and Gaia based catalogues a detailed comparison would be needed, which is out of the scope of the work presented here.
A complete sample is relevant for determining absolute cluster counts, while an incomplete representative sample can still be used to determine relative counts and study cluster formation and evolution.  
As it has been shown that the MWSC sample is magnitude-limited \citep[see][]{mwscint,mwscage,mwscmass}, construction of such a representative subsample is possible with the appropriate technique. Therefore, we prefer to use MWSC data for our observational base.

While it may be desirable to augment MWSC data using high-precision Gaia astrometry, this cannot be done from existing Gaia-based catalogues alone due to limited overlap in cluster lists, and would in practice require a full redetermination of all cluster basic parameters using Gaia data. Similarly, validating the MWSC data using existing catalogues is a non-trivial task that may be complicated by selection effects from limited catalogue overlap. As such, both approaches are beyond the scope of this work.

The MWSC catalogue \citep{khea12,mwscat}, represents a heliocentric sample of Galactic star clusters built using astrometry of PPMXL \citep[][]{ppmxl} in combination with near-infrared photometry from 2MASS \citep[][]{cat2mass}.
{It contains} basic parameters for 3061 open clusters \citep{mwscat}, 202 of which are newly discovered as part of the MWSC survey \citep{mwscnew,mwscnew2}.
Investigation by \citet{mwscint} determines that the MWSC sample is magnitude-limited.
\citet{mwscmass} extend the list of basic cluster parameters by the cluster tidal masses.

As the MWSC sample of open clusters can be characterised as magnitude-limited, it is necessary to address the issue of in-sample bias {due to variable completeness. For this purpose, we followed here the approach of `magnitude-dependent completeness limits' used by \citet{mwscage,mwscmass}, which requires the assumption of a uniform distribution of clusters in the Galactic plane in the solar neighbourhood. Each cluster was assigned an individual completeness distance
\be
\hat{d}_{xy} = p - q I(M_{K_\mrm{S}}),
\ee
depending on its integrated magnitude $I(M_{K_\mrm{S}})$ in the $K_\mrm{S}$ passband. Here we adopted the values $p = 0.36 \,\mrm{kpc}$ and $q = 0.54\, \mrm{kpc \, mag^{-1}}$ from \citet{mwscmass}.
Only clusters which have a heliocentric distance projected to the Galactic plane $d_{xy}$ smaller than their individual magnitude-dependent completeness limit, that is, which fall inside their completeness area 
\be
S(M_{K_\mrm{S}}) = \pi \hat{d}_{xy}^2,
\ee
contribute to the {`representative'} subsample of clusters. We find that of the 3061 MWSC open clusters, $N_\mrm{Cl}=2227$ fall inside their magnitude-dependent completeness limits.

{This completeness correction results in a sample with a single formal global completeness fraction $f_\mrm{G}$ applying to all cluster counts, such that the true cluster population is larger than the observed one by a factor $f_\mrm{G}^{-1}$. We omit this factor over the course of our following analysis as we implicitly deal only with observed cluster counts and derived quantities, and note that in-sample determination of $f_\mrm{G}$ is, in general, not possible. }

We consider now the distribution of clusters in the plane spanned by cluster ages and masses.
For ease and brevity of notation, we denote in the following
\be
\tau = \log t / \mrm{yr} \,\, \mrm{and} \,\, \mu = \log m/\msun
\ee
as shorthand for logarithmic ages and masses. {Here we use $\log$ to denote the decadic logarithm and $\ln$ for the natural logarithm.}

We define the CAMF as a number surface density of clusters per interval in age and mass
in linear and logarithmic space by
\be
\varsigma (m,t) \mrm{d}t \, \mrm{d}m =
\tilde{\varsigma} (\mu,\tau) \mrm{d}\tau \, \mrm{d}\mu ,
\ee
resulting in
\be
\label{eq:camf_logscaling}
\tilde{\varsigma} (\mu,\tau) = \frac{m\, t}{(\log \mrm{e})^2}\varsigma(m,t).
\ee

From the observed clusters, we can construct the CAMF for some finite logarithmic age step $\Delta_k \tau$ and mass step $\Delta_l \mu$ by summing over the contributions $S_i^{-1}$ of the $\Delta_{k,l}N$ clusters inside the age-mass interval to the cluster surface density:
\be
\label{eq:camf_binned}
\tilde{\varsigma}_{k,l} = \frac{1}{\Delta_k \tau \Delta_l \mu} \sum\limits_{i=1}^{\Delta_{k,l}N} \frac{1}{S_i}.
\ee

If we now consider that the cluster ages and masses are only known up to a certain accuracy, we can extend the CAMF to include these uncertainties by considering for a given age-mass interval the probability $P_{i,k,l}$ of a given cluster to have age and mass falling into the intervals $\Delta_k \tau$ and $\Delta_l \mu$, respectively. Equation (\ref{eq:camf_binned}) then becomes a sum over all $N_\mrm{Cl}$ clusters as
\be
\label{eq:camf_error}
\hat{\varsigma}_{k,l} = \frac{1}{\Delta_k \tau \Delta_l \mu} \sum\limits_{i=1}^{N_\mrm{Cl}} \frac{P_{i,k,l}}{S_i}.
\ee

The $P_{i,k,l}$ were computed by assuming a Gaussian error in logarithmic age-mass space. We define the Gaussian function
\be
G(x,\delta) = \frac{1}{\sqrt{2\pi}\delta} \exp \left(-\frac{x^2}{2\delta^2}\right),
\ee
and write
\be
P_{i,k,l} = \int_{\Delta_k \tau} G(\tau - \tau_i , \delta_{\tau,i} ) \mrm{d}\tau \int_{\Delta_l\mu} G(\mu - \mu_i , \delta_{\mu,i} ) \mrm{d}\mu,
\ee
where $\tau_i$ and $\delta_{\tau,i}$ are the cluster's logarithmic age and its uncertainty, and $\mu_i$ and $\delta_{\mu,i}$ are the cluster's logarithmic mass and its uncertainty.

Uncertainties in the tidal masses are given in the MWSC catalogue for all clusters for which tidal masses were determined. However, only a small subset of clusters has explicitly given age uncertainties. We referred to \citet[Table 6]{mwscat} for upper and lower bounds on the uncertainty of cluster age determination and used estimated uncertainties of $\delta_\tau = 0.1$ for clusters with age $\tau > 8.2$ and $\delta_\tau = 0.14$ for clusters with age $\tau \leq 8.2$.

{We note that this approach is conceptually very similar to that of kernel density estimation, where some kernel function is used to estimate an underlying smooth density function from a set of discrete data points. Here, however, kernels of different size were used for the different clusters, and directly represented the uncertainties of cluster age-mass determination.

This `error-smoothed' CAMF possesses two advantages in representing the underlying `true' CAMF compared to the binned CAMF of Eq.~(\ref{eq:camf_binned}). Firstly, the issue of sparseness and the effects of binning, which particularly affect the low-density edges of the CAMF, are significantly reduced. Secondly, the strongly varying uncertainties of the properties of individual clusters are taken into account.}

Our completeness-corrected sample of MWSC clusters spans a range of tidal masses $\mu=0$--4.4 and ages $\tau=6$--9.8.
For the sake of our analysis, we excluded 57 young clusters with $\tau < 6.5$. The reasoning for this is twofold: Firstly, the age determination was done using Padova isochrones, for which the lower age limit is $\tau = 6.6$, and secondly clusters younger than 3~Myr may still be in the gas-embedded, star-forming phase, which we do not consider here. This leaves us with a sample of 2170 clusters.

\begin{figure}
    \centering
    \includegraphics[width=1.0\hsize]{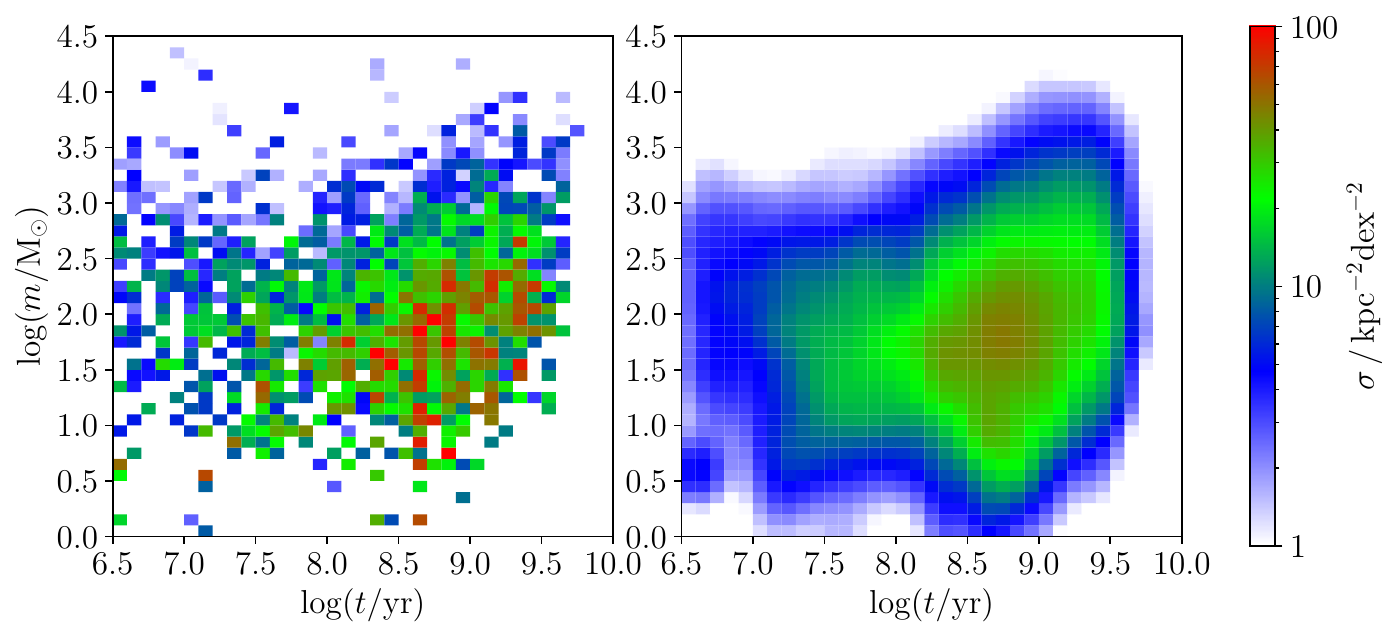}
    \caption{Observed logarithmic cluster age-mass distribution constructed for step sizes $\Delta \tau = \Delta\mu = 0.1 \,\mrm{dex}$. The left panel contains the CAMF constructed just from cluster ages and masses (see Eq.~(\ref{eq:camf_binned})) and the right panel contains the CAMF according to Eq.~(\ref{eq:camf_error}), taking into account uncertainties in age and mass.}
    \label{fig:observed_camf}
\end{figure} 

In Fig.~\ref{fig:observed_camf} the CAMFs constructed according to Eqs.~(\ref{eq:camf_binned}) and (\ref{eq:camf_error}) are plotted for logarithmic step sizes $\Delta \tau = 0.1$ and $ \Delta\mu = 0.1$.
We can see already some features of the CAMF that bear relevance for modelling.
{
\begin{itemize}
    \item The upper-limit of observed cluster age at $\tau = 9.7$ appears relatively independent of cluster mass. This observation suggests that the cluster lifetime is only weakly dependent on the cluster mass for cluster lifetimes close to this age limit.
\item The high-mass contours change with cluster age in a specific way.
The logarithmic age bins cause an increasing number of clusters per bin at fixed mass as bin sizes in linear age grow, leading to mass contours rising with age (cf.~Eq.~(\ref{eq:camf_logscaling})). Cluster mass loss shifts the contours to lower masses as age increases, counteracting the effect of logarithmic binning. For ages $\tau > 7.8$, the high-mass edge rises with age, which means the effect of logarithmic binning is stronger than that of cluster mass loss. The almost constant high-mass edge for $\tau < 7.8$ suggests that cluster mass evolution is more rapid for these young clusters.
(An alternate visualisation can be found in App.~\ref{app:vr}.)
\end{itemize}
}

\section{Cluster formation and evolution model}\label{sec:model}

We used a model for the CAMF that is based directly on the model presented in \citet{mwscmass}, which itself is an extension of an earlier model used in \citet{mwscage} to model the cluster age distribution.

We define the infinitesimal cluster surface density at age $t$ and mass $m$
\be
\label{eq:camf_differential}
\sigma(m,t) \mrm{d}m \mrm{d}t = \Psi(t) \mrm{d}t \, f(M) \mrm{d}M
\ee
via the product of the cluster formation rate (CFR) $\Psi(t)$ and the cluster initial mass function (CIMF) $f(M)$, where $M$ is the initial cluster mass directly after formation.
The CFR gives the rate at which clusters form as a function of time in terms of cluster surface density per unit time and the CIMF gives the initial mass distribution of newly formed clusters, here normalised to unity.

The initial mass $M$ of a cluster and its current mass $m$ at time $t$ are related by a bound-mass function $m(M,t)$ which models cluster mass evolution and is implicitly assumed to be invertible, such that at a given age, exactly one initial mass corresponds to any given $m$.
The mass evolution of real clusters is a deeply stochastic process \citep[driven for instance by IMF sampling, stellar evolution and complex gravitational dynamics; cf.~e.g.][]{2021A&A...655A..71W, 2008MNRAS.383..897E} which depends not just on the initial mass but on the specific dynamical state. Therefore, this model of mass evolution must be understood on the population level as the average mass evolution of an ensemble of individual clusters with fixed initial mass $M$.
This mass evolution model also accounts for the destruction of clusters, as clusters with an evolved bound mass of 0 have become fully unbound and dissolved, and no longer contribute to the CAMF.

Consequently, the modelled age-mass distribution is given by
\be
\label{eq:camf_model}
\sigma(m,t) =  \Psi(t) f(M) \frac{\partial M}{\partial m}.
\ee
We note that $\sigma(m,t)$ is computed using the inverse $M(m,t)$ of the mass evolution function $m(M,t)$.
As this inverse may in general not have an analytic expression, it is useful to keep in mind the relation
\be
\left. \frac{\partial M(m,t)}{\partial m} \right. = \left. \left( \frac{\partial m (M,t)}{\partial M} \right)^{-1}  \right.
\ee
for the mass derivative ${\partial M}/{\partial m}$ appearing in Eq.~(\ref{eq:camf_model}).

Analogously to the observed CAMF, we can, using some model $\delta_\tau (\mu,\tau)$ and $\delta_\mu (\mu,\tau)$ for the typical logarithmic error in age and mass respectively, define
\be 
\tilde{\sigma} (\mu,\tau) = \frac{m \,t}{(\log \mrm{e})^2} \sigma(m,t)
\ee
and
\be
\label{eq:camf_model_error}
\begin{split}
\hat{\sigma} (\mu,\tau) = \int\!\!\!\int &\tilde{\sigma} (\mu^\prime,\tau^\prime) G(\tau - \tau^\prime , \delta_\tau (\mu^\prime ,\tau^\prime ) ) \\&\times G(\mu - \mu^\prime , \delta_\mu (\mu^\prime ,\tau^\prime ))  \mrm{d}\tau^\prime \mrm{d}\mu^\prime 
\end{split} 
\ee
as a CAMF including uncertainty of age and mass determinations.
{For the continuous model distribution, the summation of Gaussian kernels over a set of clusters is represented by a convolution with a Gaussian function.}

Values of the model CAMF for an age-mass bin can be computed via
\be
\tilde{\sigma}_{k,l} = \int_{\Delta_k \tau} \int_{\Delta_l \mu} \tilde{\sigma} (\mu,\tau) \mrm{d}\mu \mrm{d}\tau
\ee
and analogously for $\hat{\sigma} (\mu , \tau )$.

In the following, we briefly describe the input functions we used.
Table \ref{tab:base_model} contains the set of parameters corresponding to the model presented by \citet{mwscmass}, which we call here the `base model'.

\subsection{Cluster formation rate}
\label{sec:model_cfr}

The {CFR} $\Psi (t)$ gives the number of newly formed clusters per squared kpc of Galactic disc surface and per Myr.
We used a constant CFR
\be
\Psi (t) = \beta,
\ee
as this should be sufficient for our relatively simple model and provide a good baseline for later comparison with more elaborate functions based for example on models for the star formation rate.

This parameter $\beta$ was chosen such that the modelled total cluster surface density $\Sigma$ was equal to the observed total surface density $\sum\limits_{i=1}^{N_\mrm{Cl}} S_i^{-1}$.
{The physical intuition behind this is that the total cluster content corresponds to the CFR by integration over time, taking the dissolution of clusters into account via their mass evolution.}
The total cluster surface density
\be
\Sigma = \int \sigma(m,t) \mrm{d}m \mrm{d}t
\ee
was computed by integrating the cluster age-mass distribution for integration bounds $\log t / \mrm{yr} = 6.5$--10 and $\log m / \msun = 0$--4.5 corresponding to the observed age-mass range of open star clusters in our sample. 

\subsection{Cluster initial mass function}
\label{sec:model_cimf}

The CIMF gives the mass distribution of clusters at initial age $t=0$.
We used a CIMF that is informed by the shape of the CMF for young clusters with ages up to 20 Myr \citep[as in][Fig.~9]{mwscmass}. It takes the shape of a two-slope broken power law with a smooth transition region and allows for an exponential, Schechter-type cut-off.
Our CIMF, defined as the {fraction} of clusters $\mrm{d}N$ in initial mass interval $\mrm{d}M$, is given by
\be
\label{eq:CIMF}
f(M) = \frac{\mrm{d}N}{\mrm{d}M} = k \left(\frac{M}{M_\ast} \right)^{- (x_1 + 1)} \left[ 1 + \left(\frac{M}{M_\ast} \right)^s \right]^{\frac{x_1 - x_2}{s}} \exp \left( - \frac{M}{m_\mrm{S}} \right)
\ee
with low-mass power-law index $x_1$ for $M < M_\ast$ and high-mass power-law index $x_2$ for $M > M_\ast$.
The exponent $s$ controls the `sharpness' of the transition between the power laws. A larger value of $s$ results in a less extended transition region around $M_\ast$ and vice versa for lower values of $s$.
Finally, $m_\mrm{S}$ is the Schechter cut-off mass above which the CIMF drops exponentially.
The CIMF was normalised to unity by the normalisation constant $k$ on the initial mass interval with lower limit $m_\mrm{lower} = 2 \,\msun$ and no upper limit.

We can express the logarithmic initial mass distribution in terms of the CIMF as
\be
\frac{\mrm{d}N}{\mrm{d}\log M} =  \frac{M}{\log \mrm{e}} f(M).
\ee

\subsection{Cluster bound-mass function}
\label{sec:model_mass_evo}

Our bound-mass function is given as the product of three different mass loss contributions
\be
\label{eq:mMt}
m(M,t) = n_\mrm{L} (t)\, n_\mrm{V} (t)\, n_\mrm{sec} (M,t)\, M ,
\ee
where $n_\mrm{L} (t)$ is the mass loss factor from stellar evolution, $n_\mrm{V} (t)$ gives the bound mass fraction from violent relaxation, and $n_\mrm{sec} (M,t)$ is the bound mass fraction from secular dynamical evolution.

The mass loss factor from stellar evolution depends on the details of stellar evolution as well as the stellar initial mass function. Following \citet{lamea10}, we approximated the remaining stellar mass fraction $n_\mrm{L} (t)$ by a third order polynomial in logarithmic time
\be
n_\mrm{L} (t) = \sum\limits_{i=0}^3 p_i \left[ \log(t / \mrm{Myr} ) \right]^i,
\ee
with the parameters $p_0,\, \ldots,\, p_3$ depending on the initial mass function and metallicity of the stellar population.
{The polynomial fits of \citet{lamea10} were done using a \cite{2001MNRAS.322..231K} IMF from 0.01 to 100$\msun$ and metallicities ranging from $Z=0.0004$ to 0.02.
\citet{mwscmass} used the values for $Z=0.008$, listed in Table \ref{tab:base_model}.
}

The violent relaxation phase that occurs in infant clusters after gas expulsion has been investigated in detail by \citet{shukirea17,shukirea18,shukirgaliyev2019,shukirea21} using N-body models. They find that the fraction of stars $n_\mrm{b}$ remaining bound to a cluster after virialisation depends strongly on the global star formation efficiency, but not on the initial mass and weakly on the initial size of the cluster.
This motivates us to formulate the bound mass fraction from violent relaxation $n_\mrm{V} (t)$ as a function of time only, without dependence on the initial cluster mass $M$.
We defined
\be
n_\mrm{V} (t) = n_\mrm{b} + (1 - n_\mrm{b}) \cosh^{-1} \left( \frac{t}{t_\mrm{V}} \right)
\ee
with bound fraction $n_\mrm{b}$ and violent relaxation timescale $t_\mrm{V}$.

We note that some star clusters do not survive gas expulsion and rapidly dissolve in this supervirial phase.
Cluster survivability is linked to star forming efficiency \citep{shukirea17,shukirea18}, density profile \citep{shukirea21}, primordial mass segregation \citep{2009ApJ...698..615V,2014MNRAS.444.3699H,2017A&A...600A..49B}, initial stellar mass function \citep{2020ApJ...904...43H}, and crucially the dynamics of gas expulsion itself \citep{2013MNRAS.428.1303S,2019MNRAS.488.3406Z}.
We did not model this early rapid disruption of clusters due to the low number of clusters observed at young ages, which makes estimating the fraction of clusters affected difficult and limits the impact on the CAMF.

The final factor $n_\mrm{sec} (M,t)$ gives the bound mass fraction from secular evolution of a star cluster in the Galactic tidal field. This mass loss characterised by slow evaporation and eventual dissolution must depend on the initial cluster mass in order to recover a dependence of the cluster lifetime on the initial cluster mass.
Our approach here mirrors that of \cite{lamgi06,lamea}, with both the mass loss rate and the cluster lifetime being characterised by power laws in initial mass. However, we did not consider different effects separately and thus did not fix parameters a priori.

We defined the secular bound fraction as
\be
\label{eq:n_secular}
n_\mrm{sec}(M,t) = \left( 1 - \frac{t}{t_\mrm{Cl}(M)} \right)^{\frac{1}{1-a_1}},
\ee
where $a_1 < 1$ and $t_\mrm{Cl} (M)$ the cluster lifetime, such that the secular evolution follows the power law
\be
\frac{\mrm{d}n_\mrm{sec}}{\mrm{d}t}(M,t) = - \frac{1}{(1 - a_1)t_\mrm{Cl}(M)} n_\mrm{sec}^{a_1} (M,t)
\ee
{for $t <  t_\mrm{Cl}(M)$. The implicit convention here is $n_\mrm{sec}(M,t) = 0$ for $t \geq t_\mrm{Cl}(M)$.}

We used the lifetime-mass relation
\be
\label{eq:ltmr}
t_\mrm{Cl}(M) = \frac{1}{c (1 - a_1)} \left( \frac{M}{\msun} \right)^{1-a_1} \left[ 1 + \left( \frac{M}{M_\mrm{br}} \right)^{a_2} \right]^{-1}.
\ee
This lifetime follows a broken two-slope power-law with index $1 - a_1$ for low masses $M \ll M_\mrm{br}$ and index $1 - a_1 - a_2$ for high masses $M \gg M_\mrm{br}$. The constant $c$ governs the overall rate of secular mass loss.
We note that this lifetime is parametrised slightly differently than in \citet{mwscmass}. There, $c$ was scaled to the lifetime at $M_\mrm{br}$, while $c$ was here scaled to the lifetimes of low-mass clusters $M \ll M_\mrm{br}$. As such, the value of $c$ given in Table \ref{tab:base_model} differs from the one of \citet{mwscmass} by a factor of $\left( M_\mrm{br} / \msun \right)^{1-a_1}$.

\begin{table} 
\caption {Parameters of the base model}\label{tab:base_model}
\centerline{\small
\begin{tabular}{l c}
\hline 
\hline
\vgap
Parameter & Value\\
\vgap
\hline
\vgap
$\beta$ & 0.81 kpc$^{-2}$ Myr$^{-1}$ \\ 
\vgap
\hline
\vgap
$k$ & $1.5\times 10^{-4}\,\msun^{-1}$ \\ 
$M_\ast$ & 1000 $\msun$\\ 
$s$ & 2.4 \\ 
$x_1$ & 0 \\ 
$x_2$ & 1.2 \\
$m_\mrm{S}$ & 85000 $\msun$ \\ 
\vgap
\hline
$p_0$ & 1.0078 \\
$p_1$ & $-0.07456$ \\
$p_2$ & $-0.02002$ \\
$p_3$ & 0.00340 \\
\vgap
\hline
$n_\mrm{b}$ & 0.1 \\
$t_\mrm{v}$ & 5 Myr \\
\vgap
\hline
$a_1$ & $-0.2$ \\
$a_2$ & 0.9 \\
$c$ & 6.34 Myr$^{-1}$ \\
$M_\mrm{br}$ & 5000 $\msun$\\
\vgap
\hline
\end{tabular}}
\end{table}

\section{Methods}\label{sec:methods}
\subsection{Error model} \label{sec:errormodel}

In order to compute a modelled CAMF that includes smoothing by age and mass errors, we need to model the measurement errors of cluster age and mass. We did this using functions $\delta_\tau (\mu,\tau)$ and $\delta_\mu (\mu , \tau )$ as introduced in Sect.~\ref{sec:model}.
As the age errors we used are relatively small and mostly the same across all ages, we used simply
\be
\delta_\tau (\mu ,\tau ) = 0.1,
\ee
as this is consistent with the uncertainties claimed by \citet{mwscat}.
For the mass errors, we considered correlations of the logarithmic cluster age and mass with the logarithmic mass error and found that the error is largely independent of the mass, but does depend on age.

As the mass errors have a large variation between clusters of similar ages and masses, we constructed mean and median errors for windows of 0.1~dex in age and mass. Using these `local' estimates of typical mass errors, we arrived at the linear relation
\be
\delta_\mu (\mu,\tau ) = 0.05 \tau - 0.025,
\ee
which provides a good description of the typical mass errors.

\begin{figure}
    \centering
    \includegraphics[width=1.0\hsize]{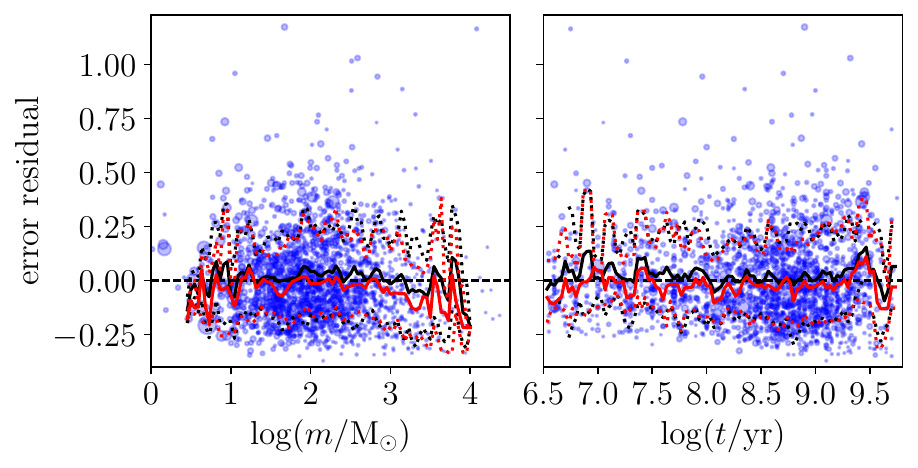}
    \caption{Error residuals $\delta_{\mu,i} - \delta_\mu (\mu,\tau)$ (circles) as functions of cluster mass (left panel) and age (right panel). The size of the circles is proportional to the clusters' contributions to the total surface density. Mean and median of the residuals (solid red and black lines, respectively) were computed using a sliding window of width 0.1~dex. Dotted lines in red and black denote the corresponding intervals enclosing one standard deviation and the 16th to 84th percentile, respectively. We note that a single outlying cluster with $\delta_\mu > 2.5$ lies beyond the $y$-axis range and is not shown.}
    \label{fig:errormodel}
\end{figure} 

The error residuals, defined here as $\delta_{\mu,i} - \delta_\mu (\mu,\tau)$, are plotted in Fig.~\ref{fig:errormodel}. The good agreement between the typical errors and the modelled errors indicates that there is little to gain by introducing higher order terms in $\mu$ or $\tau$ to the modelled error.
We note that while the error model appears to be no longer valid at the extremes of the observed ranges for cluster ages and masses (i.e.~for $\mu < 0.3$, $\mu > 3.9$ or $\tau > 9.6$), there are few (43 in total) clusters in these ranges. As such, this possible systematic deviation, if not an artefact of poor statistics, is not significant compared to the large scatter of cluster mass errors.

{When using this error model to compute an error-smoothed model CAMF following Eq.~(\ref{eq:camf_model_error}) to compare to the observed error-smoothed CAMF constructed according to Eq.~(\ref{eq:camf_error}), we applied a factor of $\sqrt{2}$ to the modelled errors. This was to account for the fact that in addition to the explicit application of the uncertainties to the CAMF, the observed values of cluster ages and masses are already scattered from their true values on a scale estimated by their given uncertainties.}

\subsection{$\chi^2$-statistic} \label{sec:chi}

The most straight-forward approach to quantifying the difference between our model and the observed cluster sample is to consider the difference between the corresponding {error-smoothed} CAMFs. We write the $\chi^2$-statistic

\be
\label{eq:chi-squared}
\chi^2 = \sum\limits_{k,l}^{N_\mrm{bin}} \frac{\left( \hat{\varsigma}_{k,l} - \hat{\sigma}_{k,l} \right)^2}{\varrho \left( \hat{\sigma}_{k,l} + \epsilon \bar{\varsigma} \right)},
\ee
where the surface density $\varrho = 1\mrm{kpc^{-2} dex^{-2}}$ scales the variance of the modelled CAMF and the constant offset $\epsilon \bar{\varsigma}$ avoids the singularity for $\hat{\sigma}_{k,l}=0$. Here 
\be
\bar{\varsigma} = \frac{1}{N_\mrm{bin}} \sum\limits_{k,l}^{N_\mrm{bin}} \hat{\varsigma}_{k,l}
\ee
is the mean observed surface density per age-mass bin.
Using this small offset is equivalent to comparing $\hat{\varsigma} + \epsilon \bar{\varsigma}$ and $\hat{\sigma} + \epsilon \bar{\varsigma}$ instead of $\hat{\varsigma}$ and $\hat{\sigma}$ directly.

\subsection{Kullback-Leibler divergence} \label{sec:KLD}

Our second approach to judge the correspondence of the model to the observed CAMF was based on the Kullback-Leibler divergence (KLD).
For two discrete probability distributions $p,\, q$, the KLD
\be
D_\mrm{KL}( p \parallel q ) = \sum\limits_i p_i \ln \frac{p_i}{q_i}
\ee
gives a measure of the difference between distributions. We note that the KLD is semidefinite positive, but not symmetric as $D_\mrm{KL}(p \parallel q) \neq D_\mrm{KL}(q \parallel p)$. Typically, $p$ is a true or observed distribution, while $q$ is a model or approximation thereof. {In this picture, the KLD gives a measure of the amount of information lost when using the model to approximate the data.}

We can convert our observed and modelled {error-smoothed} CAMFs into probability distributions by normalising them to unity, letting us write
\be
\label{eq:kld}
D_\mrm{KL}(\hat{\varsigma} + \epsilon \bar{\varsigma} \parallel \hat{\sigma} + \epsilon \bar{\sigma} ) = \frac{1}{N_\mrm{bin}} \sum\limits_{k,l}^{N_\mrm{bin}} \frac{\hat{\varsigma}_{k,l} + \epsilon \bar{\varsigma}}{(1 + \epsilon) \bar{\varsigma}} \ln \left( \frac{\hat{\varsigma}_{k,l} + \epsilon \bar{\varsigma}}{\hat{\sigma}_{k,l} + \epsilon \bar{\sigma}} \right) - \ln \left( \frac{ \bar{\varsigma}}{\bar{\sigma}} \right),
\ee
where $\bar{\sigma} = \frac{1}{N_\mrm{bin}} \sum\limits_{k,l}^{N_\mrm{bin}} \hat{\sigma}_{k,l}$, for the KLD between observed and modelled CAMFs including measurement uncertainties. As with the $\chi^2$-statistic, we included an artificial constant offset regulated by $\epsilon$ which avoids singularities that arise when the modelled CAMF goes to zero. 

\subsection{Expected likelihood}\label{sec:likelihood}

In the previous section, we argue that by normalising the binned CAMF to unity, we can interpret it as a discrete probability distribution, where the value of the normalised CAMF for a given bin corresponds to the probability of a random cluster having age and mass within the 2D interval of that age-mass bin.
We extend this concept now to the continuous CAMF: the normalised model CAMF, $\sigma (m,t)/\Sigma$, gives the probability mass distribution for clusters forming and evolving according to the model. That is, for a given model, the probability of a single cluster having age and mass in the interval $[t,t+\mrm{d}t]\times [m,m+\mrm{d}m]$ is given by 
\be
P_\mrm{M}(m,t) \,\mrm{d}m \,\mrm{d}t = \frac{\sigma (m,t)}{\Sigma} \mrm{d}m \,\mrm{d}t.
\ee

To obtain from the true age and mass the observed age and mass of a cluster, we need to model the probability mass function for a cluster's observed age and mass conditioned on the true values: $P_\mrm{O\vert M}( m_\mrm{obs},t_\mrm{obs}\vert m,t)$. 
We argue that we can approximate this function using a Gaussian centred on the true values with widths in age and mass given by the measurement uncertainties.
Indeed this holds if the measurements are unbiased and the errors are Gaussian. This is the same assumption we have already used in Sect.~\ref{sec:data} when constructing the observed CAMF including age-mass uncertainties (see Eq.~(\ref{eq:camf_error}) and following).
We have then
\be
P_\mrm{O\vert M}( m_\mrm{obs},t_\mrm{obs}\vert m,t) \approx \frac{m\, t}{(\log \mrm{e})^2} G(\tau - \tau_\mrm{obs} , \delta_{\tau,\mrm{obs}} ) G(\mu - \mu_\mrm{obs} , \delta_{\mu,\mrm{obs}} ),
\ee
where again $\mu_\mrm{obs} = \log m_\mrm{obs}/\msun$, and $\tau_\mrm{obs} = \log t_\mrm{obs} / \mrm{yr}$. The factor $m\, t / (\log \mrm{e})^2$ arises from transforming the Gaussian error in logarithmic age-mass space into linear age-mass space.

Further, the probability mass function for an observed cluster of given age and mass, implicitly conditioned on the model, is
\be
P_\mrm{O}(m_\mrm{obs},t_\mrm{obs}) = \int\!\!\!\int P_\mrm{O\vert M}(m_\mrm{obs},t_\mrm{obs} \vert m,t) P_\mrm{M}(m,t) \,\mrm{d}m \,\mrm{d}t,
\ee
and we can then approximate
\be
\label{eq:pobs_integral}
\begin{split}
P_\mrm{O}(m_\mrm{obs},t_\mrm{obs}) \approx \int\!\!\!\int &\frac{m\, t}{(\log \mrm{e})^2} G(\tau - \tau_\mrm{obs} , \delta_{\tau,\mrm{obs}} ) \\&\times G(\mu - \mu_\mrm{obs} , \delta_{\mu,\mrm{obs}} ) P_\mrm{M}(m,t) \,\mrm{d}m \,\mrm{d}t
\end{split}
\ee
for all observed clusters.

Now we construct an estimator of the expected value of the log-likelihood for a cluster from our sample 
\be
\label{eq:ell}
E[\ln P_\mrm{O}] = \sum\limits_{i = 1}^{N_\mrm{Cl}} n_i \ln P_\mrm{O}(m_i,t_i),
\ee
where
\be
n_i = \frac{1}{ S_i \Sigma},
\ee
with $\Sigma =  \sum\limits_{k = 1}^{N_\mrm{Cl}} S_k^{-1}$. $n_i$ represents the fraction of clusters that have the same properties as the $i$-th cluster of our sample, using the same argument by which the binned normalised CAMF represents a discrete probability distribution of cluster age-mass intervals.

We used this expected log-likelihood as a measure of model fitness by the following argument:
Given a representative sample $\mathcal{S}$ of $N$ clusters and a prior probability of the model $P_\mrm{prior}$, the model likelihood given the sample is, by Bayes' theorem,
\be
\ln P(\mrm{model} \vert \mathcal{S}) = \ln P_\mrm{prior} - \ln P(\mathcal{S}) + \sum\limits_{i=1}^N \ln P_\mrm{O}(m_i,t_i),
\ee
where $P(\mathcal{S})$ is the (unknown) probability of the sample, which does not depend on the choice of the model.
Since for $N \rightarrow \infty$
\be
\frac{1}{N}\sum\limits_{i=1}^N \ln P_\mrm{O}(m_i,t_i ) \rightarrow E[\ln P_\mrm{O}(m,t)],
\ee
maximising the expected log-likelihood of a single cluster also maximises the model likelihood for a given sample of sufficiently large size. In this sense, use of the expected log-likelihood is equivalent to a maximum likelihood approach.

The reason that we did not use the likelihood of the observed cluster sample directly lies in the fact that our sample features individual completeness limits. As in Eq.~(\ref{eq:ell}), we can construct relative weights $n_i$ for the individual cluster log-likelihoods using the completeness areas, however it is not readily evident what effective sample size $N_\mrm{eff} = \sum\limits_{i = 1}^{N_\mrm{Cl}} n_i$ should be used to construct the total sample likelihood. As such, we used $\sum\limits_{i = 1}^{N_\mrm{Cl}} n_i = 1$ to estimate only the expected log-likelihood for a single cluster.

We note that when evaluating Eq.~(\ref{eq:pobs_integral}) numerically, we limited the integration domain to the 3$\sigma$-intervals of the Gaussians.

\subsection{Parameter constraints and priors}\label{sec:parameters}

From the list of parameters as given in Table \ref{tab:base_model}, we varied the parameter vector
\begin{align*}
\boldsymbol{\theta} = \lbrace & \log M_\ast / \msun ,\, s ,\, x_1 ,\, x_2 ,\, \log m_\mrm{S} / \msun ,\,\\&  \log c / \mrm{yr}^{-1} ,\, a_1 ,\, a_2 ,\,  \log M_\mrm{br} / \msun , \\& \log n_\mrm{b} ,\, \log t_\mrm{V} / \mrm{yr} \rbrace,
\end{align*}
which gave us a total of $N_\mrm{par} = 11$ free parameters.
In particular, since we fit model to data in the logarithmic age-mass plane, we varied parameters that represent times and masses such as $M_\ast,\, m_\mrm{S} ,\, M_\mrm{br}$ and $t_\mrm{V}$ in logarithmic space directly, likewise $c$, which has dimension of inverse time, and $n_\mrm{b}$, which is a mass fraction.

Regarding the remaining parameters, $k$ normalises the CIMF and was thus fixed by the other CIMF parameters. Since $k$ has inverse mass dimension, we give in the context of parameter fitting also its value in logarithmic space.
The parameter $\beta$ was fixed given all other model parameters by matching the modelled total cluster surface density with the observed one (see for this Sect.~\ref{sec:model_cfr}).
The parameters of $\mu_\mrm{L} (t)$ remained fixed to the values from \citet{lamea10} for $Z=0.008$ given in Table \ref{tab:base_model}. This was because the mass loss from stellar evolution gives a comparatively small contribution, such that varying it gives little benefit. Moreover, $\mu_\mrm{L} (t)$ depends on underlying physical assumptions regarding stellar evolution and the stellar IMF that are not easily translated into constraints for the parameters $p_0,\, p_1 ,\, p_2$ and $p_3$. As such, in the context of parameter fitting, one approach would be to instead vary the underlying parameters such as the metallicity and compute the corresponding values of the $p_i$ by interpolating from tabulated values.

\begin{table} 
\caption {Parameter priors and constraints}\label{tab:par_constraints}
\centerline{\small
\begin{tabular}{l c c c}
\hline 
\hline
\vgap
Parameter & Prior & Range & Constraints\\
\vgap
\hline
\vgap 
$\log M_\ast / \msun$ & $2.5\pm 1$ & $[0,5]$&\\ 
$s$ && $[0,20]$ & \multirow{2}{*}{ $x_1 < x_2$} \\ 
$x_1$ & $0\pm0.6$ & $[-2,2]$ & \multirow{2}{*}{ $\frac{x_2-x_1}{s} \in (0.2,5)$} \\ 
$x_2$ & $1\pm0.3$ & $[0,2]$ & \\
$\log m_\mrm{S} /\msun$ && $[3,6]$ & \\ 
\vgap
\hline
$\log c / \mrm{yr^{-1}}$ && $[-11,11]$ & \multirow{2}{*}{ $\tau_\mrm{Cl} (1\msun) < 10 $}\\
$a_1$ && $[-1.2,1]$ & \multirow{2}{*}{ $ \tau_\mrm{Cl} (10^6\msun) > 9.5 $ } \\
$a_2$ && $[0,2.1]$ & \multirow{2}{*}{ $1-a_1-a_2 > 0.1$ }\\
$M_\mrm{br}$ && $[-1,7]$ & \\
\vgap
\hline
$\log n_\mrm{b}$ &$-0.6\pm0.4$& $[-2,0]$ & \\
$\log t_\mrm{v} / \mrm{yr}$ &$7.2\pm0.7$& $[6,8]$ & \\
\vgap
\hline
\end{tabular}}
\end{table}

The remaining parameters were constrained in consideration of three aspects: The definition of the model functions, the computability of the model CAMF and the physicality of the model.
The first aspect refers to the fact that certain parameters are required to have certain properties, such as, for instance, non-negativity, by the definition of their corresponding model functions. As such, we required non-negativity for $x_1,s,$ and $a_2$, as well as $x_1 < x_2$, $a_1 < 1$ and $\log n_\mrm{b} < 0$.
The second aspect mainly refers to the requirement for {the} lifetime-mass relation to be monotonic such that $m(M,t)$ can be inverted to compute $M$ for given $m,t$. We required $1 - a_1 - a_2 > 0.1$ instead of simple positiveness in order to avoid poor numerical conditioning. Similarly, we imposed finite bounds for $s$ in order to avoid undesirable behaviour of the CIMF in the limits $s \rightarrow 0$ and ${(x_2 - x_1)}/{s} \rightarrow 0$.
The third aspect refers to the observational and physical constraints we impose on the model from prior knowledge. We know that some clusters reach ages $\tau > 9.5$, but others dissolve within the observed range of ages, which gives us constraints for the cluster lifetimes. The existence of clusters of a given mass places constraints on the Schechter cut-off mass. Prior work on N-body dynamics gives limits on the timeframe during which violent relaxation will take place \citep[see][]{shukirea17,shukirea18,shukirgaliyev2019,shukirea21}.

In Table \ref{tab:par_constraints}, the ranges to which each parameter was confined are given in the third column, and the more complex constraints involving multiple parameters are listed in the fourth column.

In the second column of Table \ref{tab:par_constraints}, Gaussian priors with mean $\mu_\mrm{G}$ and standard deviation $\sigma_\mrm{G}$ are given in the form $\mu_\mrm{G}\pm \sigma_\mrm{G}$ for some parameters. These priors represent information that could easily be extracted from the data without performing a full fit in {eleven}-dimensional parameter space.

As discussed in \citet{mwscmass} (see in particular Table 3 therein), the high-mass power-law slope of the CMF is close to $x_2 = 1$ independent of age, while the low-mass slope changes significantly as the cluster population evolves, being shallow initially and steeper for higher cluster ages. Extrapolating from this, we adopted priors for $x_1$ and $x_2$ that reflect a high-mass slope close to one and a more uncertain, but likely shallow low-mass slope.

Regarding the parameters of violent relaxation, we considered the first two moments of the cluster mass as a function of age, in particular the mean and the standard deviation of the cluster mass. Within the context of our model, the bound mass fraction $n_\mrm{V}(t)$ applied as a factor equally to both of these moments. Assuming that violent relaxation dominates the early mass evolution, we could thus extract information on $n_\mrm{b}$ and $t_\mrm{V}$ from them.

We identified a transition in the cluster masses consistent with our expectations of violent relaxation
in the age range of $\tau = 7$--7.5. As such, we compared masses for age ranges $\tau = 6.5$--7 and $\tau = 7.5$--7.8 to estimate $n_\mrm{b}$. We found $\log n_\mrm{b} = -0.42\pm 0.13$ from the mean cluster mass and $\log n_\mrm{b} = -0.53\pm 0.07$ from the standard deviation.
These figures, however, likely underestimate the effect of violent relaxation as they neglect mass loss during the first 10~Myr of cluster evolution. The prior we adopted for the bound mass fraction is thus slightly lower. The transition in the mass moments at ages $\tau = 7$--7.5 informed the prior for the relaxation timescale $t_\mrm{V}$.

Finally, we considered again the CMF for Initial-Age clusters ($\tau = 6.4$--7.3 as per \citet{mwscmass}). Here the transition between power-law slopes occurs around $\mu = 2.3$. As this age group overlaps with the period of violent relaxation, we can infer that probably $\log M_\ast / \msun > 2.3$ and $\log M_\ast / \msun + \log n_\mrm{b} < 2.3$. As such, the prior for $n_\mrm{b}$ informed our prior for $M_\ast$.

For the remaining parameters, we did not find Gaussian priors. Instead, we used flat priors which are uniform within the parameter constraints.

\subsection{Markov chain Monte Carlo sampling}\label{sec:mcmc}

The parameter fitting was then done using the standard Bayesian approach where both the likelihood $L (\boldsymbol{\theta}) = P(\mrm{sample} \vert \boldsymbol{\theta})$ of observing our cluster sample given a model with parameter vector $\boldsymbol{\theta}$ and the prior probability $ P(\boldsymbol{\theta})$ of the parameters are combined into the posterior probability $\mathcal{P} (\boldsymbol{\theta})$ according to Bayes' theorem.
Logarithmically, we can write
\be
\ln \frac{\mathcal{P}(\boldsymbol{\theta})}{\mathcal{P}_0} = \ln \frac{L(\boldsymbol{\theta})}{L_0}  + \ln \frac{P(\boldsymbol{\theta})}{P_0},
\ee
for some normalising constants $\mathcal{P}_0, \, L_0$, and $P_0$ for posterior, likelihood, and prior, respectively.
Explicitly knowing these normalising constants is not necessary to sample the posterior probability distribution, which was done using a MCMC approach. By constructing a sufficiently large and representative sample of the posterior, we can estimate parameter uncertainties and correlations as well as find best fit values of the parameters by finding parameter vectors which maximise the posterior probability.

In detail, this approach was varied here in how the likelihood $L (\boldsymbol{\theta})$ is estimated.
For the $\chi^2$-statistic (see Eq.~(\ref{eq:chi-squared})), we used
\be
\ln \frac{L}{L_0} = - \chi^2_\mrm{red},
\ee
where $\chi^2_\mrm{red} = \frac{\chi^2}{n_\mrm{dof}}$ is the reduced $\chi^2$-statistic for $n_\mrm{dof} = n_\mrm{bins} - n_\mrm{pars} - 1$ degrees of freedom. We used an offset with $\epsilon = 0.005$.

For the KLD (see Eq.~(\ref{eq:kld})), we used
\be
\ln \frac{L}{L_0} = - \kappa_\mrm{KLD} D_\mrm{KL}(\hat{\varsigma} + \epsilon \bar{\varsigma} \parallel \hat{\sigma} + \epsilon \bar{\sigma} ) 
\ee
with scaling constant $\kappa_\mrm{KLD} = 20$ and offset parameter $\epsilon =10^{-4}$.

Similarly, we used
\be
\ln \frac{L}{L_0} =  \kappa_\mrm{EL} E[\ln P(m,t \vert \boldsymbol{\theta})] 
\ee
with scaling constant $\kappa_\mrm{EL} = 5$ for the expected log-likelihood (see Eq.~(\ref{eq:ell})), representing the model by its parameter vector.

{Both scaling constants were chosen such that the likelihood varied across similar-sized ranges for all three fit statistics.}

We note that in our approach, the likelihood $L (\boldsymbol{\theta})$ is not formally the probability of the observed data given a model with parameters $\boldsymbol{\theta}$. Instead, it represents a degree of plausibility of the data arising from such a model based on how well they agree as measured by the fit statistic used.
In this sense, the posterior $\mathcal{P} (\boldsymbol{\theta})$ does not directly assign probabilities to different models, but is instead used to gauge what kinds of models can plausibly explain our cluster sample. A MCMC sample of $\mathcal{P} (\boldsymbol{\theta})$ is thus a set of plausible model parameters where models that appear to be more plausible are more frequent within the sample.

{Nonetheless, the fit statistics can be related to the model log-likelihood (by a constant factor and a normalisation offset) under certain additional assumptions.
$-\chi^2$ describes the likelihood asymptotically if the number of clusters contributing to each age-mass bin is Poisson-distributed. For large cluster numbers per bin, the Poisson distribution is approximated by a Gaussian distribution, and the terms of the $\chi^2$-statistic each become proportional to the exponent in the Gaussian corresponding to the bins.
Following our discussion in Sect.~\ref{sec:likelihood}, $E[\ln P]$ describes the log-likelihood for a cluster sample with uniform completeness across its entire range of ages and masses.
The relation between KLD and log-likelihood is more subtle. It can be shown generally that a model maximising the log-likelihood will minimise the KLD, which implies that close to the maximum likelihood parameters, log-likelihood and KLD differ (up to third order) only by a constant factor and offset.}

We used the Python \verb|emcee| module \citep{emcee} for MCMC sampling, which uses an implementation of the affine-invariant ensemble sampler described by \citet{goodman2010}. This ensemble sampler runs a number $N_\mrm{walker} > N_\mrm{par}$ of samplers in parallel, using the space spanned by them to sample the high-dimensional parameter space more efficiently.
In terms of run setup, we broadly followed the advice of \citet{emcee} as well as the \verb|emcee| documentation. We used $N_\mrm{walker} = 52$ for our 11-dimensional parameter space, running the chain in 13 parallel CPU processes using the \verb|emcee| module's multiprocessing support. Convergence of the chain was tested using the estimated integrated autocorrelation time $\hat{t}_{\boldsymbol{\theta \theta}}$ of the parameters, as computed using the \verb|emcee| module. It was further used in taking a more representative set of samples from the final chain by discarding the first $\max (2\hat{t}_{\boldsymbol{\theta \theta}})$ samples as burn-in and taking only every $\min (\hat{t}_{\boldsymbol{\theta \theta}}/2)$-th sample of the remaining chain to ensure independence of the samples. The resulting set of parameter combinations was then taken to be a sufficiently large and representative sample of the posterior.

\begin{table}
    \caption{Overview of MCMC runs.}
    \label{tab:mcmc_runs}
    \centerline{\small
\begin{tabular}{l c c c}
\hline 
\hline
\vgap
fit statistic & $\chi^2$ & $D_\mrm{KL}$ & $E[\ln P]$\\
\vgap
\hline
\vgap 
MCMC steps & 98000 & 105000 & 78000 \\
acceptance fraction & 0.197 & 0.197 & 0.221 \\
burn-in & 1951 & 2092 & 1548 \\
thinning & 292 & 306 & 238 \\
sample size & 17056 & 17472 & 16692 \\
\vgap
\hline
\end{tabular}}
\end{table}

The number of sampler steps taken until convergence as well as the burn-in and thinning for all three runs are listed in Table~\ref{tab:mcmc_runs}.
Another parameter we list to gauge the well-behavedness of the chain is the acceptance fraction for parameter proposals across all walkers and steps. \citet{emcee} suggest that well-behaved chains will typically have acceptance fractions of 20\% to 50\%. {The run using $E[\ln P]$ fulfils this criterion, and the runs using $\chi^2$ and $D_\mrm{KL}$ have acceptance fractions still very close to 20\%.}

\section{Results}
\label{sec:results}

\subsection{Best-fitting models}
\label{sec:best_models}

\begin{table*}
    \caption{Parameter fit results}
    \label{tab:model_par_fit}
    \centering\small
    \begin{tabular}{l c c c c c}
        \hline
        \hline
        \vgap
        \vgap
        Parameter & Base & $\chi^2$-fit & $D_\mrm{KL}$-fit & $E[\ln P]$-fit & Fit to Hunt \& Reffert sample\\
        \vgap
        \vgap
        \hline
        \vgap
        \vgap 
        $\beta ,\, \mrm{kpc^{-2} Myr^{-1}}$    & 0.81 & $0.8448_{-0.51}^{+2.1}$ & $1.015_{-0.73}^{+1.5}$ & $0.4090_{-0.21}^{+2.7}$ & $9.162_{-5.6}^{+71}$\\[0.8ex]
        $\langle M \rangle ,\, \msun$ & 505 & $95_{-44}^{+260}$ & $87_{-25}^{+380}$ & $224_{-180}^{+290}$ & $219_{-190}^{+280}$\\[0.8ex]
        CFR, $\msun \mrm{kpc^{-2} Myr^{-1}}$ & 409 & $80.2_{-20}^{+210}$ & $88.4_{-26}^{+230}$ & $91.6_{-50}^{+250}$ & $2011_{-1100}^{+3000}$\\
        \vgap
        \vgap
        \hline
        \vgap
        \vgap
        $\log k / \msun^{-1}$ & $-3.8$ & $-3.268_{-0.76}^{+1.3}$ & $-2.718_{-1.4}^{+0.57}$ & $-3.254_{-0.59}^{+1.4}$ & $-3.751_{-0.18}^{+2.7}$\\[0.8ex]
        $\log M_\ast  / \msun$ & 3 & $2.495_{-0.87}^{+0.55}$ & $1.980_{-0.19}^{+1.2}$ & $2.565_{-1.1}^{+0.45}$ & $2.765_{-1.9}^{+0.24}$\\[0.8ex]
        $s$      & 2.4 & $1.160_{-0.24}^{+4.3}$ & $1.955_{-0.93}^{+3.7}$ & $1.888_{-0.86}^{+4.1}$ & $2.159_{-1.1}^{+5.3}$\\[0.8ex]
        $x_1$    & 0   & $0.078_{-0.72}^{+0.21}$ & $0.109_{-0.80}^{+0.13}$ & $-0.062_{-0.68}^{+0.32}$ & $0.156_{-1.5}^{+0.18}$\\[0.8ex]
        $x_2$    & 1.2 & $1.073_{-0.25}^{+0.30}$ & $0.974_{-0.16}^{+0.41}$ & $1.046_{-0.20}^{+0.33}$ & $1.140_{-0.34}^{+0.38}$\\[0.8ex]
        $\log m_\mrm{S}/ \msun$ & 4.9 & $3.321_{-0.054}^{+1.7}$ & $3.950_{-0.73}^{+0.95}$ & $3.914_{-0.42}^{+1.5}$ & $4.743_{-0.86}^{+0.75}$\\
        \vgap
        \vgap
        \hline
        \vgap
        \vgap
        $\log n_\mrm{b}$     & $-1$ & $-0.525_{-0.48}^{+0.14}$ & $-0.552_{-0.45}^{+0.18}$ & $-0.521_{-0.48}^{+0.20}$ & $-0.474_{-0.67}^{+0.24}$\\[0.8ex]
        $\log t_\mrm{V} / \mrm{yr}$  & 6.7 & $7.262_{-0.66}^{+0.38}$ & $7.383_{-0.82}^{+0.23}$ & $7.006_{-0.45}^{+0.65}$ & $6.934_{-0.58}^{+0.80}$\\
        \vgap
        \vgap
        \hline
        \vgap
        \vgap
        $a_1$       & $-0.2$ & $-0.347_{-0.56}^{+0.33}$ & $-0.665_{-0.26}^{+0.66}$ & $-0.086_{-0.80}^{+0.13}$ & $0.538_{-1.2}^{-0.11}$\\[0.8ex]
        $a_2$       & 0.9  & $0.922_{-0.74}^{+0.17}$ & $1.455_{-1.3}^{-0.31}$ & $0.808_{-0.65}^{+0.21}$ & $0.067_{+0.048}^{+0.97}$\\[0.8ex]
        $\log c / \mrm{yr^{-1}}$ & $-5.2$ & $-5.647_{-1.1}^{+1.2}$ & $-5.129_{-1.6}^{+0.80}$ & $-6.196_{-0.81}^{+1.6}$ & $-6.988_{-0.20}^{+1.9}$\\[0.8ex]
        $\log M_\mrm{br} / \msun$ & 3.7 & $3.097_{-2.6}^{+2.0}$ & $2.683_{-2.1}^{+2.2}$ & $3.253_{-2.9}^{+2.1}$ & $6.026_{-5.9}^{-0.89}$\\
        \vgap
        \vgap
        \hline
    \end{tabular}
\end{table*}

\begin{figure*}
    \centering
    \includegraphics[width=18cm]{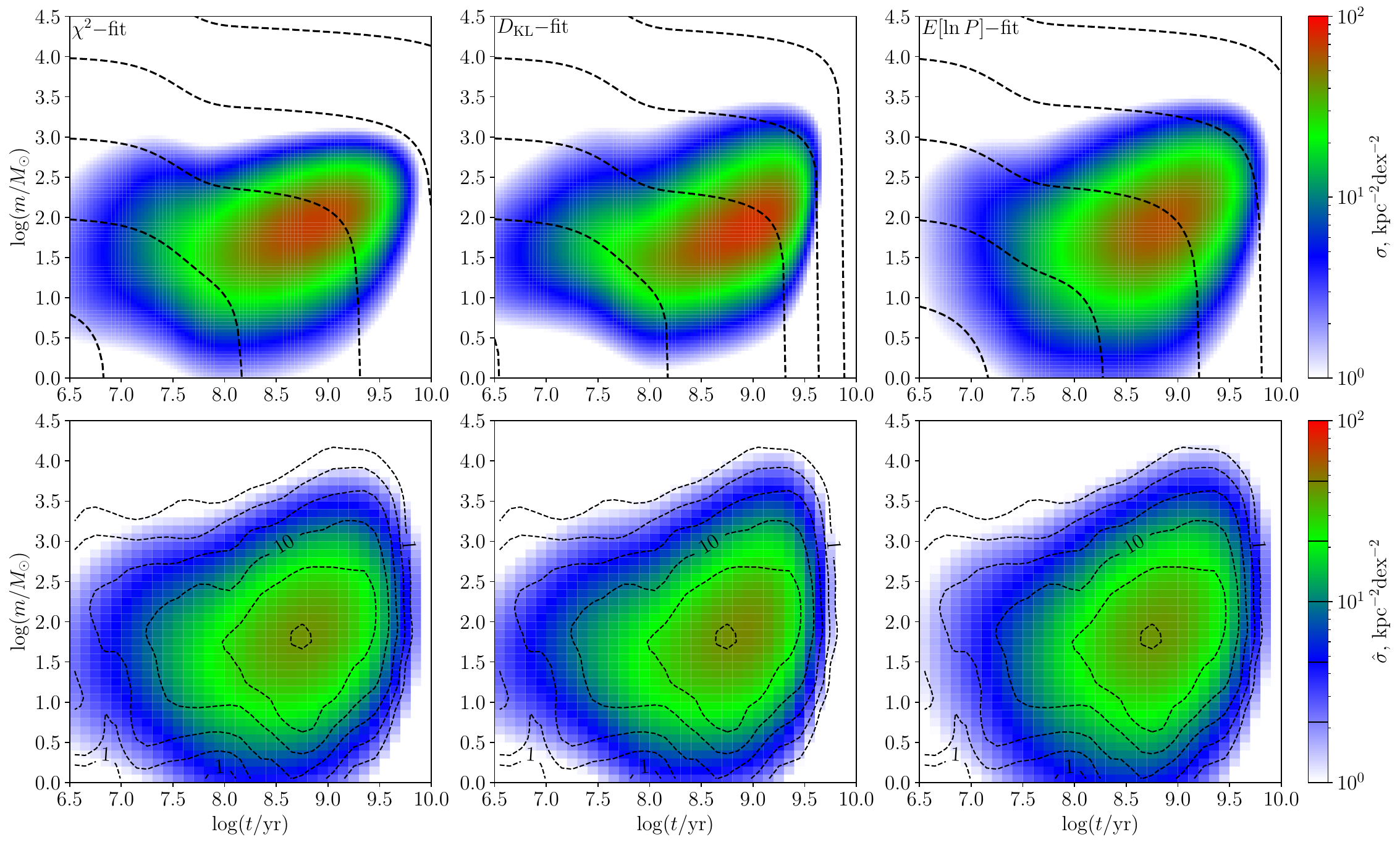}
    \caption{Modelled CAMFs following Eq.~(\ref{eq:camf_model}) for the best-fit parameters with mass loss tracks (top row) in comparison to their smoothed counterparts following Eq.~(\ref{eq:camf_model_error}) overplotted with contours of the smoothed observed CAMF (bottom row). Contour levels of the smoothed data are marked with solid lines on the colour scale bar.}
    \label{fig:camfs_bestfit}
\end{figure*}

The results of the fits are summarised in Table \ref{tab:model_par_fit}. For each MCMC run, the parameter set with highest posterior probability was chosen as the best-fitting model. The interval from the 16th to the 84th percentile was used to estimate the uncertainty of the individual parameters. This is motivated {by} that this interval, containing the central 68\% of the total probability, corresponds to the $1\sigma$-interval of a Gaussian distribution.

We note that while the best-fitting models are those which best agree with the data within each MCMC sample, they are not best-fit models in the sense of representing a local optimum of a fit statistic. In the same sense, the percentile ranges do not constitute a fitting error, but a range of values the parameters typically take for models that agree reasonably well with the data. 
Nonetheless, for brevity we refer to the MCMC runs in the following as $\chi^2$-fit, $D_\mrm{KL}$-fit, and $E[\ln P]$-fit, and to the respective best-fitting models as the $\chi^2$-fit model, the $D_\mrm{KL}$-fit model, and the $E[\ln P]$-fit model.
In principle, it is possible to use the best-fitting parameter sets as starting points to find such local optima, but as we discuss in the following, such a best-fit model may not actually be a useful improvement.

\begin{table}
    \caption{Fit statistics for the best-fitting models}
    \label{tab:fitstats_bestfit}
    \centering\small
    \begin{tabular}{l c c c c}
        \hline
        \hline
        \vgap
         Fit statistic & Base & $\chi^2$-fit & $D_\mrm{KL}$-fit & $E[\ln P]$-fit \\
        \vgap
        \hline
        \vgap 
         $\chi^2_\mrm{red}$ & 1.057 & 0.651 & 0.545 & 0.939 \\
         $D_\mrm{KL}$ &  0.0620 & 0.0407 & 0.0304 & 0.0541\\
         $E[\ln P]$ & $-2.064$ & $-2.082$ & $-2.045$ & $-2.048$ \\
        \vgap
        \hline
    \end{tabular}
\end{table}

The values of the different fit statistics are given in Table~\ref{tab:fitstats_bestfit} for the base model as well as for the different best-fitting models. As expected, the best-fitting models perform better than the base model for all statistics, {with the exception of the $\chi^2$-fit model having a lower $E[\ln P]$ value}. We note also that the $D_\mrm{KL}$-fit model outperforms (in the sense of better or equally good fit statistics) the other best-fitting models in the same manner, while the $E[\ln P]$-fit model appears to perform the worst {in terms of $\chi^2_\mrm{red}$ and $D_\mrm{KL}$}.

The CAMFs corresponding to the best-fit models are plotted in Fig.~\ref{fig:camfs_bestfit} in comparison to the observed CAMFs. 
{The top row gives the bare modelled CAMF with mass loss tracks for initial masses $\mu = 1,\,2,\,3,\,4,\,5,$ and 6. The bottom row gives the modelled CAMF including age and mass uncertainties according to our error model from Sect.~\ref{sec:errormodel} multiplied by a factor $\sqrt{2}$, together with contours of the observed CAMF smoothed using individual cluster uncertainties.

Comparing the different best-fit models with the data, we find that the $\chi^2$-fit model does not reproduce the high-mass edge of the observed CAMF due to a low Schechter cut-off scale (${\log m_\mrm{S}/ \msun = 3.3}$) which suppresses high mass clusters with $\mu > 3$. Even accounting for observational uncertainties, the observed old high-mass clusters with $\tau > 8.5,\,\mu > 3.8$ are not represented in the model. In the $D_\mrm{KL}$-fit model and the $E[\ln P]$-fit model, a Schechter cut-off scale that is four times higher allows for the existence of clusters with $\mu > 3.5$ and a better reproduction of the observed high-mass edge.

Further, the sharp age cut-off at $\tau = 9.7$ is apparently difficult to reproduce using our model, due to both its sharpness and high uniformity across cluster masses. Only the $D_\mrm{KL}$-fit model reproduces such a sharp age cut-off, because it is the only model which features a shallow increase of the cluster lifetime at the high mass end ($t_\mrm{Cl}\propto M^{1-a_1-a_2}= M^{0.21}$).}

Other features, such as the high-density region of clusters with intermediate ages and masses as well as the low-mass limit are more consistently reproduced across the models considered.
We notice also that our models all produce an overabundance of low-mass clusters ($\mu < 1$) at intermediate ages $\tau = 7.7$--8.5.

\subsection{Comparison of fit statistics}
\label{sec:fitstat_comp}

\begin{table}
    \caption{Kolmogorov-Smirnov statistic of the marginalised posterior against the prior for parameters with Gaussian priors}
    \label{tab:ks_priors}
    \centering\small
    \begin{tabular}{l c c c c}
        \hline
        \hline
        \vgap
        Parameter & $\chi^2$-fit & $D_\mrm{KL}$-fit & $E[\ln P]$-fit \\
        \vgap
        \hline
        \vgap
        $M_\ast$ & 0.1326 & 0.0831 & 0.1465 \\
        $x_1$ & 0.1669 & 0.1862 & 0.1791 \\
        $x_2$ & 0.1354 & 0.1439 & 0.1627 \\
        $n_\mrm{b}$ & 0.0938 & 0.0833 & 0.0296 \\
        $t_\mrm{V}$ & 0.0257 & 0.0258 & 0.0064 \\
        \vgap
        \hline
    \end{tabular}
\end{table}

Considering the parameter posteriors in Table \ref{tab:model_par_fit}, we find some common results that appear independent of the fit statistic used.

Firstly, the CIMF slopes behave as expected, with $x_2 \approx 1$ and a shallower $x_1 < 0.3$. We note that while we can infer an upper limit to the low-mass slope, it is more difficult to find a lower limit above the slope of $-0.7$ observed in the evolved CMFs at higher ages. The reason for this lack of constraint may be the limited statistics from observations of clusters with initial masses much smaller than $M_\ast$.

Secondly, we find that the parameters of violent relaxation are not significantly constrained beyond their priors. 
To quantify this, we considered the Kolmogorov-Smirnov statistic (KS-statistic) of the marginalised posteriors for the respective priors, that is, the maximum of the absolute difference between the cumulative distribution functions of the marginalised posterior and the prior. As we see in Table~\ref{tab:ks_priors}, {the KS-statistic is significantly lower for $t_\mrm{V}$ than for the parameters of the CIMF, independent of the fit statistic used. For $n_\mrm{b}$, the KS-statistic is still below all but the lowest value found for any combination of fit statistic and CIMF parameter.}
The reason for this may be that while the entire cluster sample contains implicit information on the shape of the CIMF, only young clusters with $\tau < 8$ contain information on violent relaxation, which is a smaller sample containing only 497 clusters. Moreover, this sample was already used to derive prior constraints, limiting the information that could be extracted additionally by the fitting procedure.

\subsubsection{Construction of synthetic cluster samples}
\label{sec:fitstat_sampling}

In order to better understand the differences between the fit statistics, we considered synthetic cluster samples as realisations of a given model. We constructed such a realisation by drawing a sufficient number $N$ of clusters from the model, applying random Gaussian errors according to the error model from Sect.~\ref{sec:errormodel}, and then evenly distributing them across a heliocentric disc of radius $R =\sqrt{{N}/\left({\pi\Sigma_\mrm{tot}}\right) }$ such that the cluster surface density corresponded to the observed value of $\Sigma_\mrm{tot} = 134.1\,\mrm{kpc^{-2}}$.
In analogy to the magnitude-dependent completeness limit, a completeness limit depending on the logarithm of its mass was drawn for each cluster.
These limits were drawn from a Gaussian with mean and standard deviation of the form
\be
\hat{d}_{xy} = w + v \mu,
\ee
with values $w_\mrm{mean} = 0.60\,\mrm{kpc},\,v_\mrm{mean} = 0.84\,\mrm{kpc}$ for the mean and $w_\mrm{std} = 0.43\,\mrm{kpc},\,v_\mrm{std} = 0.12\,\mrm{kpc}$ for the standard deviation obtained by performing linear regression on the completeness limits of the observed cluster sample. A minimum completeness distance of 0.2~kpc was chosen to disallow divergent contributions to the synthetic CAMF.
These completeness limits then also allowed the definition of a sufficient number of clusters such that the radius of the resulting heliocentric disc is at least as large at the greatest completeness distance. In practice, this was done by choosing $N$ such that $R \geq 6.8\,\mrm{kpc}$.

The final synthetic sample was then obtained by selecting only clusters within their completeness limits.
While this procedure produces samples with similar properties as the observed one by coarsely approximating the measurement and selection process used to arrive at a {completeness-corrected} sample, it cannot directly fix global properties of the samples such as number of clusters and total cluster surface density, which as a result vary between realisations and models.

For each of the base, $\chi^2$-fit, $D_\mrm{KL}$-fit, and $E[\ln P]$-fit models, 5000 realisations of synthetic cluster samples were drawn and the fit statistics with respect to the corresponding models computed.
While the total cluster surface density and cluster number of these realisations was not fixed, their construction causes them to reproduce the observed cluster surface density well, with a stochastic scatter of about 6~kpc$^{-2}$. The corresponding number of clusters also varies, with systematic deviations from the cluster number of the observed sample being introduced by the differing structure of the model CAMFs. In particular, 
realisations of $\chi^2$-fit model and $E[\ln P]$-fit model tend to produce fewer clusters, while realisations of the $D_\mrm{KL}$-fit model produce cluster numbers consistent with the observed sample.

\subsubsection{Fit statistic distributions for model realisations}
\label{sec:fitstat_qval}

Further, the fit-statistic distributions allowed us assign a $Q$-value to the observed value of the fit statistic. This represents the quantile at which the observation sits and gives the fraction of model realisations with a lower degree of agreement with the model as measured by the statistic, that is, a larger value for $\chi^2$ and KLD, and a smaller value for $E[ \ln P ]$. This $Q$-value gives an indication how typical the level of disagreement between model and data would be for a realisation of the model, both in terms of atypically high or low values of the statistic.
A very low $Q$-value means that it is unlikely that a random realisation of the model will disagree with the model at least as much as the data. As such, it signals that the model does not fit the data well.
Conversely, a very high $Q$-value means that it is unlikely for random realisations of the model to agree with the model better than the data. In this sense, it signals that the model may be `overfitted'.
What we refer to here as overfitting is when a model is fitted to not just a dataset's general features, but also its specific noise. Such a model will produce spurious features, which is why this is an undesirable property.
When overfitting occurs for a given combination of data and fit statistic, the data will match the model better than a typical realisation of the model itself, as a greater discrepancy is expected simply from the stochastic noise included in the forward modelling.

\begin{table*}
    \caption{Statistics for synthetic cluster samples from the best-fitting models}
    \label{tab:stats_bestfit_samples}
    \centering\small
    \begin{tabular}{l c c c c}
        \hline
        \hline
        \vgap
          & Base model & $\chi^2$-fit model & $D_\mrm{KL}$-fit model & $E[\ln P]$-fit model \\
          Statistic & \multicolumn{4}{c}{$Q$-values for the fit statistics}\\
        \vgap
        \hline
        \vgap 
         $\chi^2_\mrm{red}$ & 0.064& 0.135& 0.171& 0.113 \\
         $D_\mrm{KL}$ & 0.018 & 0.072 &0.139 & 0.065\\
         $E[\ln P]$ &  0.469&  0.055& 0.027& 0.112 \\
        \vgap
        \hline
        \vgap
        & \multicolumn{4}{c}{Global statistics of synthetic samples} \\
        \vgap
        \hline
        \vgap
        $\Sigma_\mrm{tot},\, \mrm{kpc^{-2}}$ & $134.1\pm 5.4$ & $134.1 \pm 5.5$ & $134.2\pm 5.4$ & $134.0 \pm 5.9$ \\
        $N_\mrm{Cl}$ & $2383\pm 46$ & $2103 \pm 43$ & $2140 \pm 43$ & $2126\pm 44$ \\
        \vgap
        \hline
    \end{tabular}
\end{table*}

The results of drawing synthetic cluster samples for the different models and computing their statistics are listed in Table~\ref{tab:stats_bestfit_samples}. This includes $Q$-values for the different fit statistics, mean cluster surface densities, and mean cluster numbers.

\begin{figure}
    \centering
    \includegraphics[width=1.0\hsize]{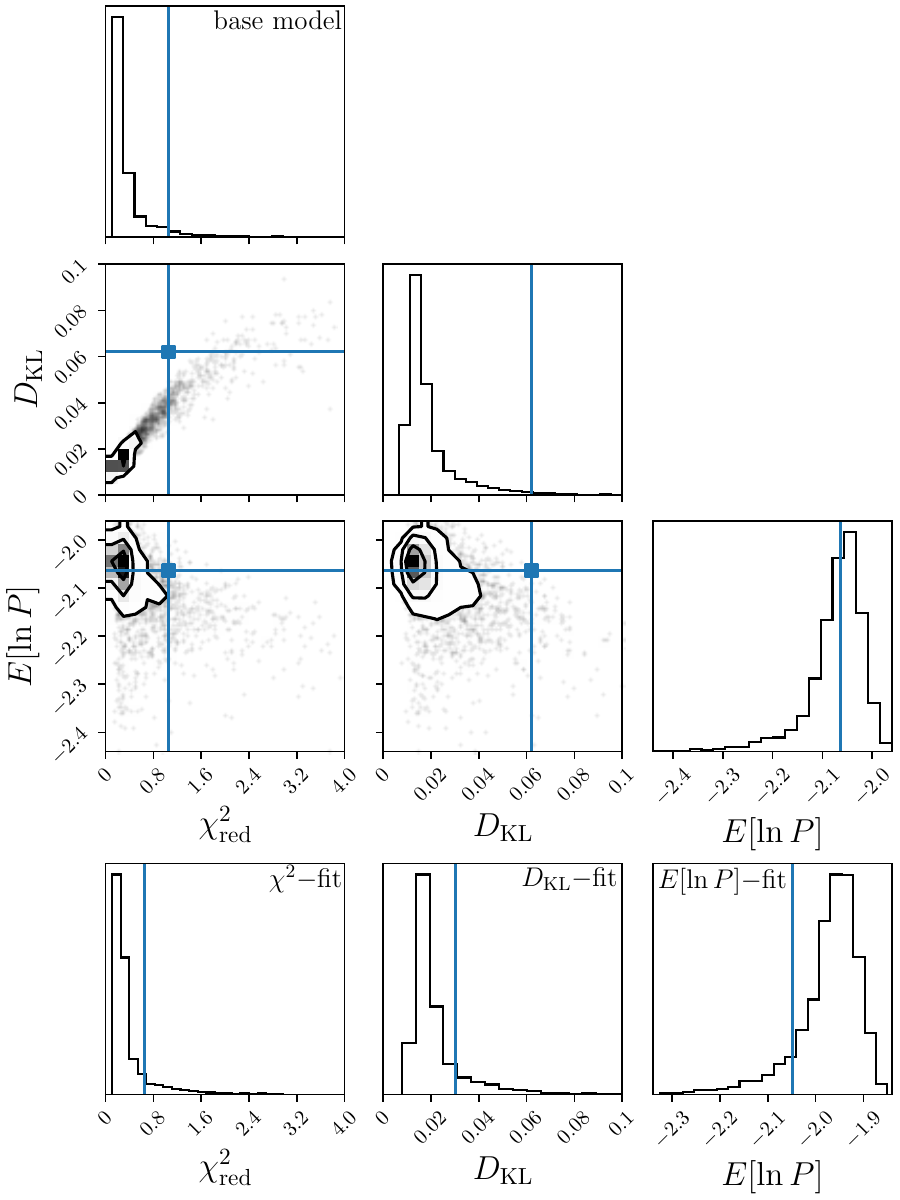}
    \caption{ Distribution of the different fit statistics for synthetic cluster samples. From left to right, the columns correspond to $\chi^2_\mrm{red}$, $D_\mrm{KL}$, and $E [\ln P ]$, respectively. The three topmost rows show the distribution for the base model, including binned 2D distributions. The contour lines enclose 11.8\%, 39.3\%, 67.5\% and 86.4\% of samples. This corresponds to levels of $0.5\sigma$, $1\sigma$, $1.5\sigma$, and $2\sigma$ for a 2D Gaussian. Samples outside the contours are drawn as individual dots. The bottom row contains histograms for samples of the $\chi^2$-fit, $D_\mrm{KL}$-fit, and $E [\ln P ]$-fit models. The blue lines mark the values of the fit statistics for the observed sample listed in Table \ref{tab:fitstats_bestfit}.}
    \label{fig:fitstat_corner}
\end{figure} 

In Fig.~\ref{fig:fitstat_corner}, the distribution of $\chi^2_\mrm{red},\, D_\mrm{KL},$ and $E[\ln P]$ for realisations of the base model is plotted, as well as the distributions of the respective fit statistics for the $\chi^2$-fit, $D_\mrm{KL}$-fit, and $E[\ln P]$-fit models. These distributions behave qualitatively very similar for the different best-fitting models, being sharply peaked with a long, one-sided tail.

{For the $\chi^2$-fit model, the observed sample has fairly low values of $D_\mrm{KL}$ and $E[\ln P]$ (each within the lowest 8\%), with the value for $\chi^2_\mrm{red}$ being more typical with a $Q$-value at 13\%.
The $\chi^2$-fit model tends to produce lower cluster numbers than observed for samples of the same total surface density, which means that the model clusters have on average lower completeness distances, corresponding to lower average masses.

In the $D_\mrm{KL}$-fit model, the value of $\chi^2_\mrm{red}$ falls within the central 68\% (corresponding to a Gaussian $1\sigma$-interval), and the value of $D_\mrm{KL}$ is still fairly typical with a $Q$-value at 14\%. However, for $E[\ln P]$, we have a very low $Q$-value of 2.7\%. As we discuss in Sect.~\ref{sec:fitstat_conc}, this may indicate the model realisations being systematically more centrally concentrated than the observed sample.
The $D_\mrm{KL}$-fit model samples also reproduce the observed cluster number well.

The $E[\ln P]$-fit model has $Q$-values close to 11\% for $\chi^2_\mrm{red}$ and $E[\ln P]$, and a lower $Q$-value of 6.5\% for $D_\mrm{KL}$. In this sense, it represents neither a particularly good nor particularly bad fit of the observation. We have already seen from the absolute values of the fit statistics for the observed sample that the $E[\ln P]$-fit model reproduces the observed CAMF better than the base model, but not as good as the $D_\mrm{KL}$-fit model, which is supported by this analysis of the model's realisations.
Similar to the $\chi^2$-fit model, it produces slightly lower cluster numbers than observed, indicating a lower average cluster mass.

Altogether, the fits appear fairly well-behaved and not close to overfitting judging by their $Q$-values. A brief comparison with the $Q$-values resulting for the base model indicates that realisations of these models are indeed closer to observations as judged through $\chi^2_\mrm{red}$ and $D_\mrm{KL}$. However for $E[\ln P]$, the base model has a high $Q$-value of 47\%, which implies that for this model, the expected likelihood cannot reliably distinguish between observations and model.}

We note that while the $E[\ln P]$-distribution for the base model shows that for some models, the expected likelihood cannot distinguish between observations and model realisations, the distribution for the $E[\ln P]$-fit model demonstrates that this does not hold in general. The low value of $E[\ln P]$ for the observed sample in the $D_\mrm{KL}$-fit model shows that there are well-fitting models that can be identified using the expected likelihood.
As such, the apparent comparative weakness of the $E[\ln P]$-fit is not necessarily the result of a deficiency of the fit statistic, but possibly of a too small value of the scaling constant $\kappa_\mrm{EL}$.

\subsubsection{General remarks and conclusions}
\label{sec:fitstat_conc}

We observe for all four models sampled a strong linear correlation between $\chi^2$-statistic and KLD, while both are more weakly correlated to the expected log-likelihood. This is seen for the base model in the 2D distributions in Fig.~\ref{fig:fitstat_corner}.
Such a linear relation between $\chi^2$-statistic and KLD when both are small is expected theoretically, since
\be
\sum\limits_i p_i \ln \frac{p_i}{q_i} =  \sum\limits_i \frac{(p_i - q_i)^2}{2 q_i} + \mathcal{O}\left((p_i - q_i)^3 \right),
\ee
which can be seen using Taylor expansion of the KLD around $p_i = q_i$.
Written in terms of $p_i$ and $q_i$, the $\chi^2$-statistic becomes proportional to $ \sum\limits_i (p_i - q_i)^2 / q_i$, and thus $\chi^2 \propto D_\mrm{KL}$ for $p_i \approx q_i$.
This relation holds for small values of $\chi^2$-statistic and KLD. For larger values of the statistics, the difference between $p_i$ and $q_i$ may become large for individual bins, leading to differing behaviours of $\chi^2$-statistic and KLD as higher-order terms of $p_i-q_i$ become relevant in the KLD.

The expected likelihood is more difficult to compare to the other two statistics as it operates without a binning of the observed clusters. This in itself may be seen as an advantage as it bypasses the problem of biases arising from the choice of the age-mass bins, but it also makes it more difficult to intuitively interpret. If we consider the expected likelihood of different representations of a model as a proxy for the case of small differences between data and model, it is clear that realisations with a high expected likelihood will have clusters from the dense regions of the CAMF overrepresented, while realisations with low expected likelihood will have those clusters underrepresented. Correspondingly, the CAMF of the model will be more densely concentrated than that of the sample if $E[\ln P]$ is small compared to a typical realisation, while in the case of a comparatively large $E[\ln P]$, the model CAMF will be more diffuse than the sample.
{This effect may be the cause for the unexpectedly high value of $E[\ln P]$ for the base model, which is less centrally concentrated than the best-fit models. This may also explain why the $D_\mrm{KL}$-fit model has a surprisingly low $Q$-value for $E[\ln P]$ despite its rather good agreement with the data as per the other fit statistics, as it may indicate that the model and its realisations are more densely concentrated than the observed sample. However, other discrepancies than underrepresentation of clusters in dense regions of the CAMF can also cause a comparatively small value of $E[\ln P]$.}

From our analysis of the fit statistics of synthetic cluster samples, it is clear that one factor which limits parameter fitting is the large impact of sampling noise on the fit statistics, which limits their ability to discriminate between models based on the observed cluster sample. This is likely because our fit statistics, as we constructed them in Sect.~\ref{sec:chi}, \ref{sec:KLD} and \ref{sec:likelihood}, do not take the Poisson noise of the cluster sample into account correctly. In our sample, faint clusters have large weight, which increases their impact on the noise.

Of our best-fitting models, the $D_\mrm{KL}$-fit model performs the best across all fit statistics and reproduces the cluster number when sampled. {It reproduces the observed sharp age limit while still allowing for high-age high-mass clusters with $\tau > 8.5,\,\mu > 3.5$ when accounting for uncertainties in mass determination. A caveat remains in that the synthetic sampling of the model shows that the value of $E[\ln P]$ for data and model is lower than for typical realisations. This may suggest that the model CAMF is more concentrated in the age-mass plane than the observed CAMF.}
As such, we took the $D_\mrm{KL}$-fit model to be the best model we found in our investigation of model parameter space. 

\subsection{Parameter correlations}

Previously, we evaluated the uncertainties of the model parameters individually based on their marginalised one-dimensional posteriors. We next considered the full eleven-dimensional posterior to derive a
better understanding of the uncertainties of the model parameters and their correlations.

As the high dimensionality of the parameter posterior distribution makes it challenging to analyse or visualise the full complexity of the posterior, we used principal component analysis in an attempt to make the posterior tractable without losing too much information.
This was done by first normalising the parameters to zero mean and unit standard deviation, and then performing an eigenvector decomposition on the covariance matrix of the posterior in normalised parameter space. The resulting eigenvectors $\mathsf{u}_i$ for indices $i=0\ldots 10$, ordered by the size of the corresponding eigenvalues, form an orthonormal basis of the normalised parameter space and are uncorrelated in the sense that $\mrm{Cov} (\mathsf{u}_i,\mathsf{u}_j) = \lambda_i \delta_{ij}$ where $\lambda_i$ is the eigenvalue corresponding to $\mathsf{u}_i$. This means that the eigenvalues represent the variance of the posterior distribution along the direction of the corresponding eigenvectors. If we consider the distribution in the eigenvector basis, then the $i$-th parameter has zero mean, standard deviation $\sqrt{\lambda_i}$ and no linear correlation to another parameter. While higher-order correlations certainly exist, this at least allows us to consider independence of these principal components as a first-order approximation. In this picture, the posterior corresponds to an ellipsoidal distribution with principal axes given by the eigenvectors $\mathsf{u}_i$.

As such, if $\lambda_i$ is smaller, then the posterior is more constrained along the axis $\mathsf{u}_i$, while it is less constrained if $\lambda_i$ is larger, with the vectors with largest eigenvalues
having the greatest contribution to the uncertainty of the parameters. Conversely, the eigenvectors with lowest eigenvalues may give us insight in the way our model is well-constrained within the posterior beyond the marginalised uncertainties of single parameters.

We note that if two or more eigenvalues are equal, their corresponding eigenvectors are not uniquely determined. Such a set of components then forms a subspace of the full parameter space in which the posterior distribution is spherical.
This is relevant for intermediate eigenvectors with eigenvalues close to unity, which may not be determined with much certainty.

In order to check that the parameters in eigenvector space are indeed less strongly correlated, we checked the marginalised posteriors in two principal components. We confirmed that the transformed posteriors are indeed significantly closer to the idealised assumption of an elliptical distribution.

\begin{figure}
    \centering
    \includegraphics[width=0.9\hsize]{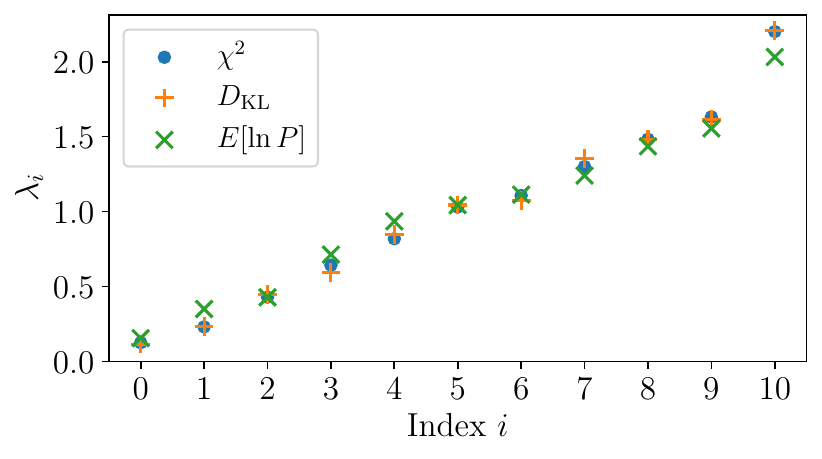}
    \caption{Eigenvalue spectra of the normalised parameter posteriors from different fit statistics.}
    \label{fig:eigenvectors}
\end{figure} 

The eigenvalue spectra for the posteriors of the different fits are shown in Fig.~\ref{fig:eigenvectors}. We can see that they agree very well between parameter fits for different fit statistics. 
In order to further check the agreement of the eigenvector spaces, we considered the angles between eigenvectors for different fits. We found that the eigenvector spaces of the posteriors from $\chi^2$-fit and $D_\mrm{KL}$-fit agree quite well, with eigenvectors with the same index corresponding to one another within angular deviations of {up to 26 degrees, and less than 9 degrees} if considering only indices $\lambda = 0,1,2,10$.
The eigenvector space of the $E [\ln P]$-fit differs more strongly from the first two eigenvector spaces. While {$\mathsf{u}_0,\, \mathsf{u}_1,\, \mathsf{u}_2$ and $\mathsf{u}_{10}$ agree within 14 degrees, the eigenvectors for intermediate eigenvalues show angular deviations between 17 and 54 degrees. In particular, $\mathsf{u}_7$ and $\mathsf{u}_8$ for the $E [\ln P]$-fit do not correspond well to $\mathsf{u}_7$ and $\mathsf{u}_8$ for the other two fits.}
This means that the posteriors agree well when it comes to the combinations of parameters subject to particularly strong or weak constraints, in particular with respect to eigenvectors $\mathsf{u}_0,\, \mathsf{u}_1,\, \mathsf{u}_2$ and $\mathsf{u}_{10}$. The intermediate eigenvectors with eigenvalues closer to unity are less certainly determined in general, so we do not expect a strong correspondence here.

Upon analysis of the impact of the different parameter eigenvectors on the model CAMF, we find that $\mathsf{u}_0$ shifts the high-age edge of the distribution along the age-axis. $\mathsf{u}_1$ and $\mathsf{u}_2$ together shift the position of the CAMF in the age-mass plane while affecting its shape only weakly. $\mathsf{u}_{10}$, which corresponds to the component of largest uncertainty, scales the CAMF along the mass-axis by shifting the low-mass edge of the distribution. 
The remaining eigenvectors affect the shape of the CAMF in more nontrivial ways and vary significantly between posteriors for different fit statistics.
We can thus link $\mathsf{u}_0$ to the observed age cut-off at $\tau = 9.7$, which apparently places the strongest constraint on the model. 
$\mathsf{u}_{10}$ can be linked to the high uncertainty of our sample's low-mass cluster content, which is caused by low-number statistics due to the faintness and corresponding difficulty of detection for these clusters.

\subsection{Compatibility of the best model with observations}

We consider now our best-fitting CAMF model, which is the $D_\mrm{KL}$-fit model following our discussion in Sect.~\ref{sec:fitstat_comp}. Its parameters can be found in the fourth column of Table \ref{tab:model_par_fit}.
We discuss first the properties of the resulting model functions and then analyse how well the model matches the observed cluster sample in more depth.

The CIMF was strongly constrained by prior considerations and is largely consistent with them. At low masses, the CIMF slope is very shallow, but at higher masses above a {transition mass of ${M_\ast = 95\msun}$,} it is close to 1. This behaviour appears consistent with the CMF constructed for clusters younger than 20~Myr in \citet{mwscmass}. {The Schechter cut-off at ${\log m_\mrm{S}/\msun = 3.950}$ lies above the majority of observed cluster masses, with ten sample clusters having higher masses.}

With a CFR of $\beta = 1.015 \mrm{kpc^{-2} Myr^{-1}}$, the mean initial cluster mass of 87$\msun$ corresponds to a contribution of 0.088$\msun\mrm{pc^{-2} Gyr^{-1}}$ of long-lived clusters to the local star formation rate (SFR).
Comparing this figure to the present-day SFR of \citet{sysjus21}, who used data from the second Gaia data release \citep{gaiadr2_18} to parametrise a model of the local Galactic disc, this corresponds to a {relative contribution of 6\%} to the present-day SFR from star formation in bound clusters.
\citet{2017MNRAS.470.1360B} used the Tycho-Gaia Astrometric Solution catalogue \citep{tgas} to reconstruct the local star forming history from the observed column densities for different types of main sequence stars. We find a {relative contribution of 5.1$\pm$1.3\%} to the present-day SFR from their work, which appears consistent with the result of \citet{sysjus21}. 

In the model, clusters undergo a violent relaxation phase during which they {lose 72\% of their member stars} on a timescale of 20~Myr. Half of this mass loss occurs between ages of 15 to 40~Myr, which means that in our model, violent relaxation happens later and more slowly than in the simulations of \citet{shukirea17,shukirea18,shukirgaliyev2019,shukirea21}.
The {bound fraction of 28\% corresponds to around 18\% global star formation efficiency} for Plummer models with centrally peaked star formation \citep{shukirea21}. However, as discussed earlier, the violent relaxation parameters are not very well constrained by our cluster sample due to the low number of clusters younger than 100~Myr.
{Accounting for all effects and integrating over cluster initial masses, half of the total initial mass is lost within 30~Myr, and 80\% becomes unbound within the first 100~Myr, making violent relaxation the dominant channel of mass loss during early cluster evolution.}

\begin{figure}
    \centering
    \includegraphics[width=0.9\hsize]{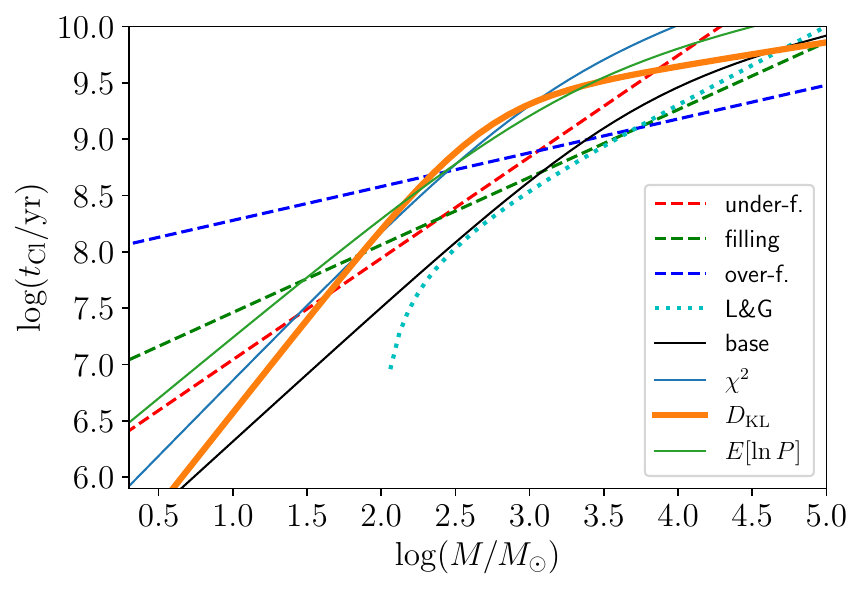}
    \caption{Comparison of cluster lifetime-mass relations. Our models are shown as solid lines, in particular the base model as well as the best $\chi^2$-fit, $D_\mrm{KL}$-fit and $E[\ln P]$-fit models. {The bold solid line corresponds to the best model.} Dashed lines correspond to \citet{mwscage} lifetimes, which follow parametrisations found by \citet{ernstea15} for Roche volume under-filling, filling and over-filling clusters. The dotted line corresponds to the mass loss model of \citet{lamgi06} for $100\msun$ remnant clusters.}
    \label{fig:ltmr}
\end{figure} 

The model's cluster lifetime-mass relation is compared to other relations from the literature in Fig.~\ref{fig:ltmr}.
We consider the relations used in \citet{mwscage}, which are based on \citet{ernstea15} half-mass time relations from N-body simulations of Roche volume under-filling, filling and over-filling clusters. We also consider the cluster mass loss model from \citet{lamgi06} for a $100\msun$ cluster remnant. The base model and the $\chi^2$-fit and $E[\ln P]$-fit models are also shown.

The best-fitting models for the $\chi^2$-fit and $E[\ln P]$-fit lie very close to the best model and follow most closely the relation for Roche volume underfilling clusters {for intermediate initial masses ${100\msun <  M < 2000\msun}$}. 
{For clusters with initial mass below $1000\msun$, the lifetime-mass relations of our models are steeper than those for filling and overfilling clusters. For $M < 200\msun$, they are also steeper than the relation for underfilling clusters.}
At high masses $M > 3000\msun$, the {best} model's lifetime-mass relation is very shallow, diverging from the other two best-fitting models {and following most closely the relation for overfilling clusters}. {Due to the high-mass restrictions imposed by the CIMF, cluster lifetimes} are effectively limited by the observed upper age limit of 5~Gyr.
Lifetimes are generally higher by a factor of about {3 to 5} compared to the base model and the model by \citet{lamgi06} {for masses ${30\msun < M < 3000\msun}$.}
The main reason for the shorter lifetimes in \citet{lamgi06} is the impact of encounters with giant molecular clouds on the cluster mass loss, which is not taken into account in our model. (To do this, an additional factor could be added to Eq. (\ref{eq:mMt}), or Eqs. (\ref{eq:n_secular}) and (\ref{eq:ltmr}) could be adjusted to include contributions from such encounters.)
These high cluster lifetimes are responsible for the comparatively low model CFR. As clusters survive longer, fewer clusters need to be formed to reach the observed total surface density integrated over all ages.

\begin{figure}
    \centering
    \includegraphics[width=0.9\hsize]{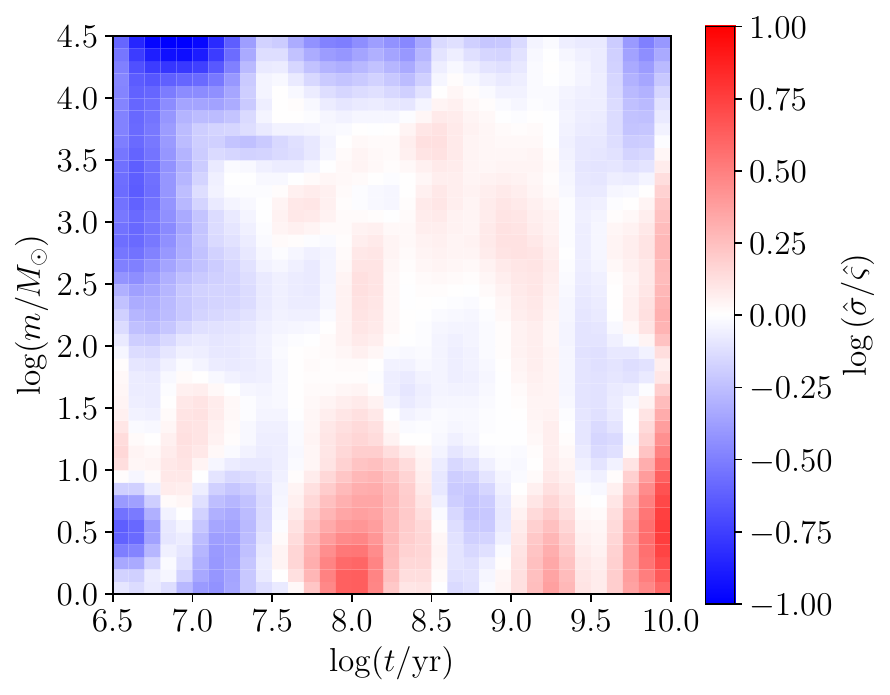}
    \caption{Log-ratio of modelled and observed CAMFs including uncertainties in cluster ages and masses, using the best-fitting model. A positive log-ratio corresponds to a modelled cluster surface density exceeding observation, and a negative log-ratio corresponds to a lower model density than observed.}
    \label{fig:camf_ratio}
\end{figure} 

In Fig.~\ref{fig:camf_ratio}, we plot the logarithm of the ratio of modelled and observed CAMF, both including uncertainties in age and mass as per Eqs.~(\ref{eq:camf_error}) and (\ref{eq:camf_model_error}). 
We observe the best agreement between CAMFs at intermediate ages and masses, where the cluster surface densities are high. 

For young clusters $\tau < 7$, the modelled cluster density is systematically lower than the observed density. We further see large disagreements between CAMFs for low masses $\mu < 1$, where the data is quite noisy due to low number statistics. 
At high masses $\mu > 3.5$, the model density is generally lower than observed. 

Except for low {masses and the outermost high-age edge where no clusters are observed}, the model tends to have {lower} densities than the observation towards the edges of the CAMF, but {higher} densities at the centre of the CAMF. {This appears to be consistent with the model CAMF being more concentrated} in the age-mass plane than the observation, as was already indicated by the $E [\ln P]$ distribution in Sect.~\ref{sec:fitstat_comp}.

In order to test the consistency of the best-fitting model with the data in more detail, we considered the CAMF in initial-mass space. We recall Eq.~(\ref{eq:camf_differential}), which implies that {in initial-mass space, the CAMF is given by the product of CFR and CIMF. By using the bound-mass function to infer initial cluster masses, we can thus reconstruct the CIMF and the CFR from the cluster sample data.}

\begin{figure}
    \centering
    \includegraphics[width=0.9\hsize]{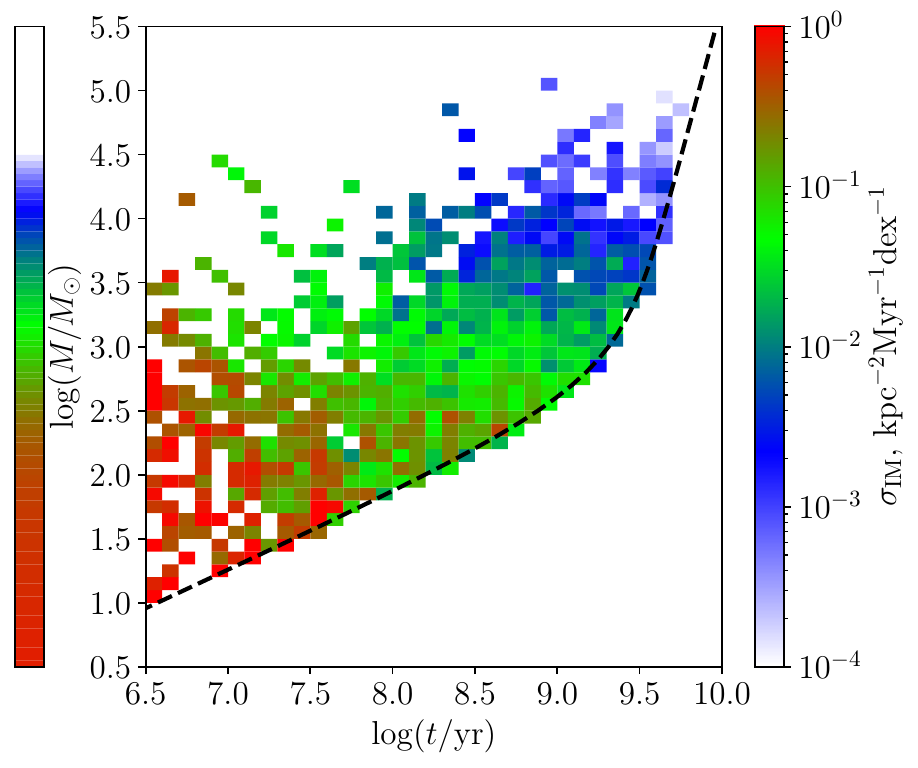}
    \caption{CAMF in initial mass space constructed using the bound-mass relation of the best-fitting model. Cluster densities were computed for linear bin widths in age and logarithmic bin widths in initial mass. The coloured bar on the left represents the product of model CIMF and CFR, which is constant with age. The dashed line represents the model lifetime-mass relation.}
    \label{fig:camf_imass}
\end{figure} 

We constructed the CAMF in initial-mass space in Fig.~\ref{fig:camf_imass}, using a linear age scale for bin normalisation in order to visualise the factorisation of the CAMF according to Eq.~(\ref{eq:camf_differential}). {As our CFR is constant and the CIMF does not depend on cluster age, this initial-mass CAMF is $\approx \beta f(M) \Delta M / \Delta\log (M/\msun )$ within a bin centred on initial mass $M$, independent of cluster age. } We see that {Fig.~\ref{fig:camf_imass}} seems compatible with the adopted constant CFR and unchanging CIMF. While we observe few higher-mass clusters with $\mu > 3.5$ at young ages $\tau < 8$, this can easily be explained by stochastic noise due to the generally lower number of clusters at younger ages.
If we want to gain further information from this CAMF, more assumptions must be made. In particular, by assuming our model CIMF to be the correct one, we can extract the CFR from the CAMF, and vice versa to extract the CIMF.

To reconstruct the CFR, we considered the clusters with ages in an infinitesimal interval $\mrm{d}t$ around age $t$.
The number of clusters formed at that age was then $\Psi (t) \mathrm{d}t$, with a surviving fraction of
\be
\label{eq:fsurv}
f_\mrm{surv} (t) = \int\limits_{M_0(t)}^\infty f(M)\mrm{d}M,
\ee
where $M_0(t)$, defined by $t_\mrm{Cl}(M_0) = t$ as the inverse of the cluster lifetime-mass relation, gives the lower limit of the initial mass required for a cluster to survive until age $t$.

The total cluster surface density at age $t$ is thus $f_\mrm{surv} (t) \Psi (t) \mathrm{d}t$, and we could then reconstruct the CFR given CIMF and cluster lifetime-mass relation by computing
\be
\hat{\Psi}_\mrm{rec} (t_i) = \frac{1}{\Delta_{t,i}f_\mrm{surv}(t_i)} \sum\limits_{t_k\in \Delta_{t,i}} S_k^{-1},
\ee
where the sum gives 
the total surface density of clusters within the age bin $\Delta_{t,i}$.

\begin{figure}
    \centering
    \includegraphics[width=0.9\hsize]{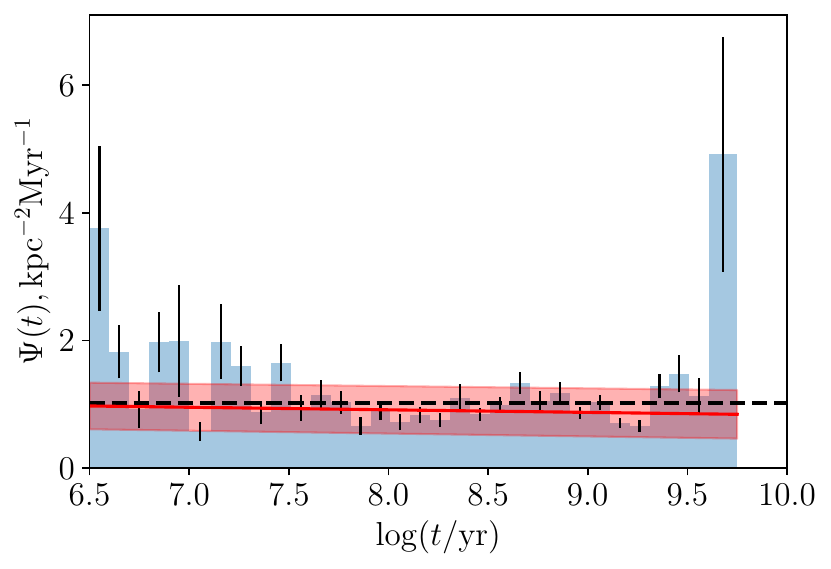}
    \caption{CFR reconstructed from the cluster sample using CIMF and lifetime-mass relation of the best-fitting model. Error bars represent uncertainties from Poisson noise. The solid red line represents a linear fit of the CFR, with the shaded area giving the uncertainty of the fit. The horizontal dashed line gives the constant CFR of the model.}
    \label{fig:cfr_reconstructed}
\end{figure} 

We reconstructed the {CFR from observations} using the CIMF and lifetime-mass relation of our model and plot it in Fig.~\ref{fig:cfr_reconstructed} together with the model CFR as well as a linear fit to the {reconstructed} CFR to check for age-dependent trends.
We note that while the model CFR in principle extends to $\tau = 10$, the {reconstructed} CFR could only be constructed up to $\tau = 9.75$, beyond which no clusters were observed.
The slope of the linear fit at $-0.040 \pm 0.005 \mrm{kpc^{-2} Myr^{-1} dex^{-1}}$, which is statistically significant, indicates that while the reconstructed CFR systematically deviates from a constant, it does so only slightly. As such, we interpret this as primarily a weakness of the model fit and not a fundamental incompatibility of the data with a constant CFR.

While the high values of the reconstructed CFR for the youngest ages may {correspond to clusters that do not survive gas expulsion and violent relaxation (as discussed in Sect.~\ref{sec:model_mass_evo}), the data for such clusters is sparse. This effect may also be explained by stochastic noise.
Similarly, the high CFR in the final age bin is likely an effect of noise. As the surviving fraction is very close to zero at the high-age edge, a small number of clusters can have a disproportionate impact on the reconstructed CFR. Still, this high CFR value could also reflect a period of enhanced star formation earlier in the Milky Way's history \citep[as e.g.~in][]{2019ApJ...887..148F,2022Natur.603..599X}.}

For reconstructing the CIMF, we considered the total cluster surface density per unit mass $n(M)$ at a given initial mass $M$.
This is given by
\be
n(M) = f(M) \int\limits_{t_\mrm{min}}^{t_\mrm{Cl}(M)} \Psi (t) \mrm{d}t = f(M) \beta (t_\mrm{Cl}(M) - t_\mrm{min})
\ee
for our model of constant CFR, where $t_\mrm{min}$ is the lower age limit of the cluster sample.
As such, we constructed the CIMF {from a cluster population} given the bound-mass function to compute cluster initial masses and a constant CFR as
\be
\hat{f}_\mrm{rec} (M_i) = \frac{1}{\beta (t_\mrm{Cl}(M_i) - t_\mrm{min})} \frac{1}{\Delta_{M,i}}\sum\limits_{M_k\in \Delta_{M,i}} \frac{1}{S_k},
\ee
where $\sum\limits_{M_k \in \Delta_{M,i}} \frac{1}{S_k}$ is the total surface density of clusters within the inital-mass bin $\Delta_{M,i}$. The logarithmic CIMF can be constructed analogously using logarithmic initial-mass bins instead.


\begin{figure}
    \centering
    \includegraphics[width=0.9\hsize]{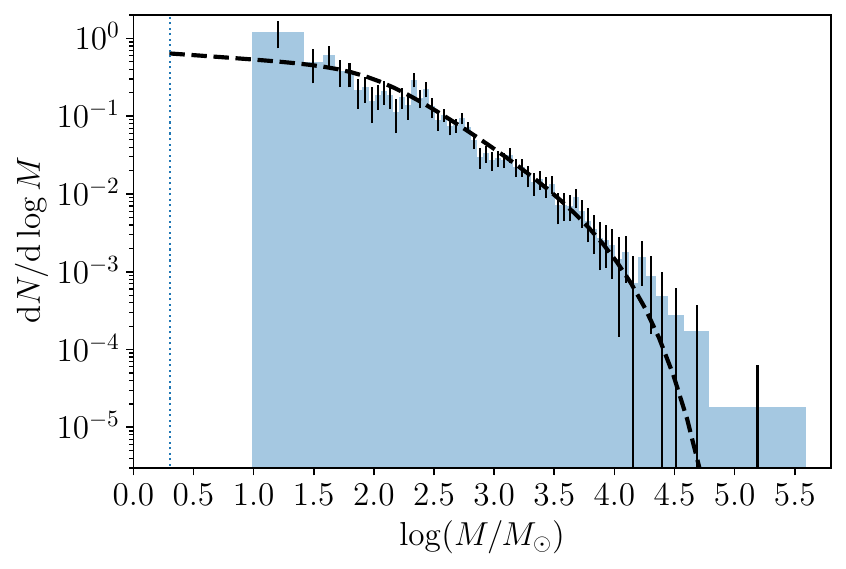}
    \caption{CIMF reconstructed from {the} cluster sample using CFR and bound-mass function of the best-fitting model. Error bars represent combined uncertainties from Poisson noise and mass determination errors. The dashed line represents the CIMF of the model. The dotted vertical line marks the lower mass limit of the model CIMF.}
    \label{fig:cimf_reconstructed}
\end{figure} 

As for the CFR, we reconstructed the CIMF using our CAMF model, in particular via bound-mass function and CFR, plotting the result in Fig.~\ref{fig:cimf_reconstructed}.
We see that the reconstructed CIMF is highly consistent with our model for intermediate masses of $\mu = 2.5$ to 4. For initial masses above $\mu = 4$, the reconstructed CIMF {does not fall off as fast as} the model CIMF, and may thus be compatible with a {higher} Schechter cutoff mass than present in the model. However, uncertainties here are relatively large and both CIMFs still mostly agree within the errors.
At lower masses, a break in the reconstructed CIMF power law is observed. Compared to the model CIMF, a sharper transition between power laws at a higher transition mass $M_\ast$ also seems permissible, possibly in combination with a slightly steeper low-mass power-law slope.
Modelled clusters with initial masses {between $m_\mrm{lower} = 2\msun$ and $10\msun$ are not observed due to their short lifetimes, which makes it difficult to assess the CIMF below the breaking point of the high-mass power-law slope.}

{This also indicates that the sample (and the best model in particular) would be consistent with a higher value of the lower limit to the initial cluster masses. However, this makes little difference to the resulting CAMF as the total cluster surface density remains fixed. When setting $m_\mrm{lower} = 10\msun$ for our best model, this resulted in changed parameter values $\log k / \msun = -2.489$ and $\beta = 0.5989\mrm{kpc^{-2}Myr^{-1}}$. All other parameter values remained unchanged, and the mean initial cluster mass increased to $\langle M \rangle = 144\msun$, such that the CFR was only slightly decreased to $86.4\msun\mrm{kpc^{-2}Myr^{-1}}$. }

Overall, we find the reconstructed CFR and CIMF to agree within their uncertainties with the corresponding model functions. We interpret this as the model being compatible with the data in a self-consistent manner, and make no further claims about the reconstructed CFR and CIMF beyond their use as a test for consistency between model and observations, as their shapes depend sensitively on the details of the cluster lifetime-mass relation and the bound mass function.

\section{Comparison to a cluster sample based on Gaia DR3}
\label{sec:gaia_data}

As we mention in Sect.~\ref{sec:data}, mass determinations of large cluster catalogues based on Gaia astrometry have so far been relatively sparse in the literature. Due to the high scientific interest in the Gaia data releases \citep[][]{tgas,gaiadr2_18,gaiaedr3_21,gaiadr3_23}, the state of the literature on Milky Way clusters using Gaia data has been rapidly evolving, so this will likely change in the near future. It seems therefore expedient to demonstrate 
that the CAMF and CAMF models can prove useful tools for analysing cluster systems 
also in the context of Gaia-based cluster catalogues.

The catalogue published by \citet{2023A&A...673A.114H} is a Gaia-based sample of 7167 clusters detected using the HDBSCAN algorithm. It contains distance measures and age determinations based on a machine learning approach using variational inference to estimate uncertainties.

During the preparation of this work, this catalogue was supplemented by \citet{hunt&reffert24a} with Jacobi masses based on apparent stellar masses and a classification of objects into bound clusters and unbound moving groups. This gives us the opportunity to demonstrate the application of our CAMF model to a different cluster sample, as well as compare our results to more recent cluster data.

As an in-depth analysis of the catalogue's completeness does not yet exist, we used the simple mass-based completeness model that is presented in \citet{hunt&reffert24a} as an approximate estimate.

Their completeness distance is defined as 
\be
R_{100\%} = \min \{ \alpha \times \ln ( M_J / \msun ) + \beta , R_\mrm{break} \},
\ee

where $M_J$ is the Jacobi mass, with parameters $\alpha = 633.1 \pm 7.3 \mrm{pc}$,
$\beta = -1582.6 \pm 39.5 \mrm{pc} $, and
$R_\mrm{break} = 2792.9 \pm 8.2 \mrm{pc}$.

Using these individual completeness limits, the construction of a representative, completeness-corrected sample proceeds analogously as for the MWSC in Sect.~\ref{sec:data}.

In addition to the distance-related correction, \citet{hunt&reffert24a} recommend a mass cut at $40\msun$, below which their classification of bound clusters and the determination of cluster masses become unreliable.
The completeness-corrected sample of the \citet{2023A&A...673A.114H, hunt&reffert24a} catalogue then uses the prescription $d_{XY} < R_{100\%}$ and $M_J > 40\msun$. In total, it contains 2481 of the 5647 open clusters listed in the catalogue.

Using the given completeness distances, ages, masses and corresponding uncertainties from the catalogue, we constructed CAMFs as we did using MWSC data.
However, the $40\msun$ lower mass cutoff enforced different mass limits than for the MWSC.

We obtained a total surface density of
\be
\Sigma_{\mu > 1.6} = 257.3 \mrm{kpc}^{-2},
\ee
contrasted to $\Sigma_{\mu > 1.6} = 86.3 \mathrm{kpc}^{-2}$ for the MWSC sample. This means that the observed cluster counts are roughly three times higher. The CAMFs both without and with consideration of individual age-mass uncertainties can be found in Fig.~\ref{fig:hr24_camfs}.

\begin{figure}
    \centering
    \includegraphics[width=1\hsize]{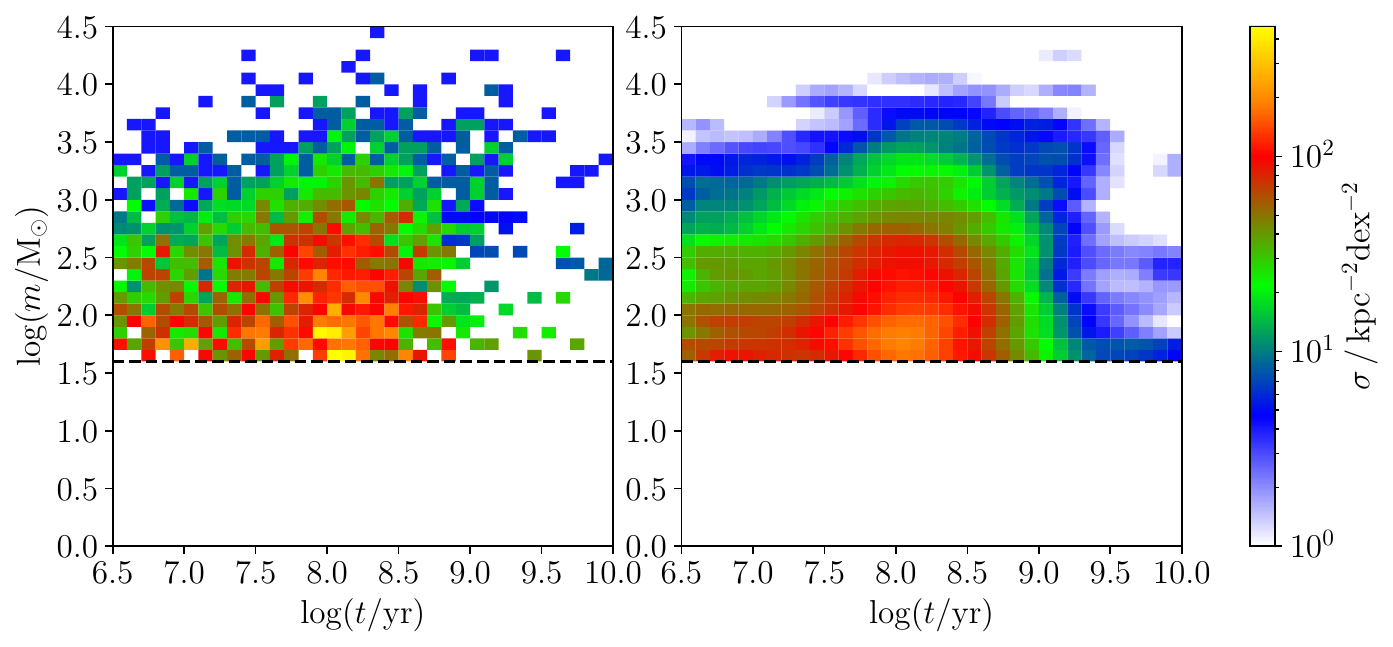}
    \caption{CAMFs derived from the Gaia-based catalogue of \citet{2023A&A...673A.114H, hunt&reffert24a}. In the left panel is the CAMF without age-mass uncertainties. In the right panel is the CAMF taking into account uncertainties in age and mass. The horizontal dashed lines mark the 40$\msun$ mass cutoff.}
    \label{fig:hr24_camfs}
\end{figure} 

We observe that like for the MWSC sample, there is evidence for a period of enhanced mass loss at early ages from the change in the high-mass contours, as well as some effective upper age limit above which cluster densities drop rapidly.
The major difference is that both enhanced mass loss and age limit, as well as the cluster density maximum, appear at earlier ages than in the MWSC sample, with a constant high mass edge for $\tau < 7.3$ and a high-age edge in the range of $\tau = $ 9--9.5. The density maximum occurs at $\tau \approx 8$ in contrast to $\tau \approx 8.7$ for the MWSC sample. 
The high-age edge is further more strongly dependent on mass as well as more diffuse. As we will discuss below, this may be related to the age-mass uncertainties and their dependence on cluster age.

In order to quantitatively compare the modelled with the observed CAMF, we again needed to take the uncertainties in age and mass into account. 
We find that age uncertainties vary weakly for ages $\tau = $ 7--9, typically remaining within $0.23\pm 0.06$~dex. For $\tau < 7$, they are slightly reduced. No clear dependence on mass was apparent. In the same age range, mass uncertainties show little systematic variation with age apart from a slight increase for $\tau < 7$, but a linear decrease of about 0.025~dex per decade in mass, with typical values at $0.07\pm 0.03$. As the cluster sample spans less than three decades in mass, this trend can account for less than 40\% of the total variance in mass uncertainties. For $\tau > 9$, age uncertainties, and to a lesser degree mass uncertainties, are higher and vary more strongly, with age uncertainties at $0.30\pm 0.19$~dex and mass uncertainties at $0.08\pm 0.04$~dex.

The increase in uncertainty coincides with the age limit and corresponding drop in the number of clusters observed. This may suggest that some of these clusters are poorly parametrised `outliers' in the sense that they are strongly observationally scattered from a different part of the true CAMF. The generally higher age uncertainties may partially explain the more diffuse high-age edge of the CAMF compared to the MWSC sample. 

We here used the simplest possible model of constant age and mass errors with
\be
\delta_\tau (\mu, \tau ) = 0.25 ,\quad \delta_\mu (\mu, \tau ) = 0.08,
\ee
while acknowledging that this is not accurate at high ages. However, at the level of this proof-of-concept analysis, we deemed it sufficient.

Using this error model in conjunction with our CAMF model, we compared our best-fitting model for MWSC data to the Gaia-based CAMF, rescaling the CFR parameter $\beta$ such that the total cluster density for $\mu > 1.6$ matches the observation.
This resulted in an alternate choice of $\beta = 3.037 \mrm{kpc^{-2} Myr^{-1}}$, about three times higher than for the MWSC sample.

As we note above, the observed CAMF generally sits at lower ages compared to the MWSC data. Likewise, the CAMF of the best-fitting model has high-age edge, density maximum and violent relaxation mass loss at higher ages compared to observation.

We used steps of 0.05~dex to find that a shift of the model to lower ages by 0.8~dex minimises the KLD between observed CAMF and error-smoothed model CAMF to $D_\mrm{KL} = 0.113$ from $D_\mrm{KL} = 0.666$ without an age shift. 
Such a shift in cluster age is equivalent to rescaling the parameters $\beta$, $c$, and $t_\mrm{V}$, which respectively govern cluster formation rate, lifetimes, and the violent relaxation timescale.

We find that while this aligns the central regions of the CAMFs, correspondence between model and data remains poor for high masses and ages.
Compared to observations, the model contains too few high-mass clusters, which are suppressed in the model by a Schechter cutoff at $\log m_\mrm{S} / \msun = 3.950$.
For $\tau > 9.5$, the model has a lack of clusters compared to the data. While the deficiency of longer-lived high-mass clusters also contributes to this, our error model underestimates the age uncertainties, leading to a lacking contribution of lower-age clusters from error smoothing.
A relative abundance of old compared to young clusters in the model relative to the data may suggest that the dissolution rate is too low in the model, which means that too many clusters survive for too long due to cluster lifetimes rising too quickly with mass.

We finally used the KLD and the above error model to perform a model fit using MCMC sampling as we did for the MWSC sample. For simplicity, we reused the same parameter constraints as they fit also this sample. For the Gaussian priors, we adopted the same means as a baseline, but used twice the standard deviations to account for the Gaussian priors being specific to the MWSC sample and avoid unduly constraining the parameter space. 

The MCMC took 221000 steps with an average acceptance rate of 0.161 until convergence. With a burn-in of 4404 samples and a thinning factor of 528, the final posterior sample contains 21320 parameter sets. The best model has a KLD of $D_\mrm{KL} = 0.034$, which represents a significant improvement compared to the age-shifted best-fitting MWSC model. The rightmost column of Table \ref{tab:model_par_fit} contains a list of this model's parameters with uncertainties derived from the posterior distribution.

As the cluster sample contains no clusters below $40\msun$, it cannot constrain the low-mass region of the CAMF and CIMF. This results in poorer constraints for the low-mass CIMF slope $x_1$. The Schechter cutoff is better constrained due to lower mass uncertainties.
In these models, $40_{-31}^{+45}$\% of clusters have initial masses below $40\msun$ and contribute to the CFR, but not the observable region of the CAMF, leading to a high uncertainty of the CFR. However, these low-mass clusters only represent a small fraction of the total cluster mass, making the CFR in terms of stellar mass formed within clusters significantly less uncertain.

The lifetime scaling parameters are no longer informed by the dissolution behaviour of clusters near the end of their life, which is encoded in the low-mass region of the CAMF. This allows for high values of $a_1$. This, in combination with a less sharp age-cutoff, results in models where the break in the cluster lifetime-mass relation can partially or even fully disappear.

\begin{figure}
    \centering
    \includegraphics[width=1\hsize]{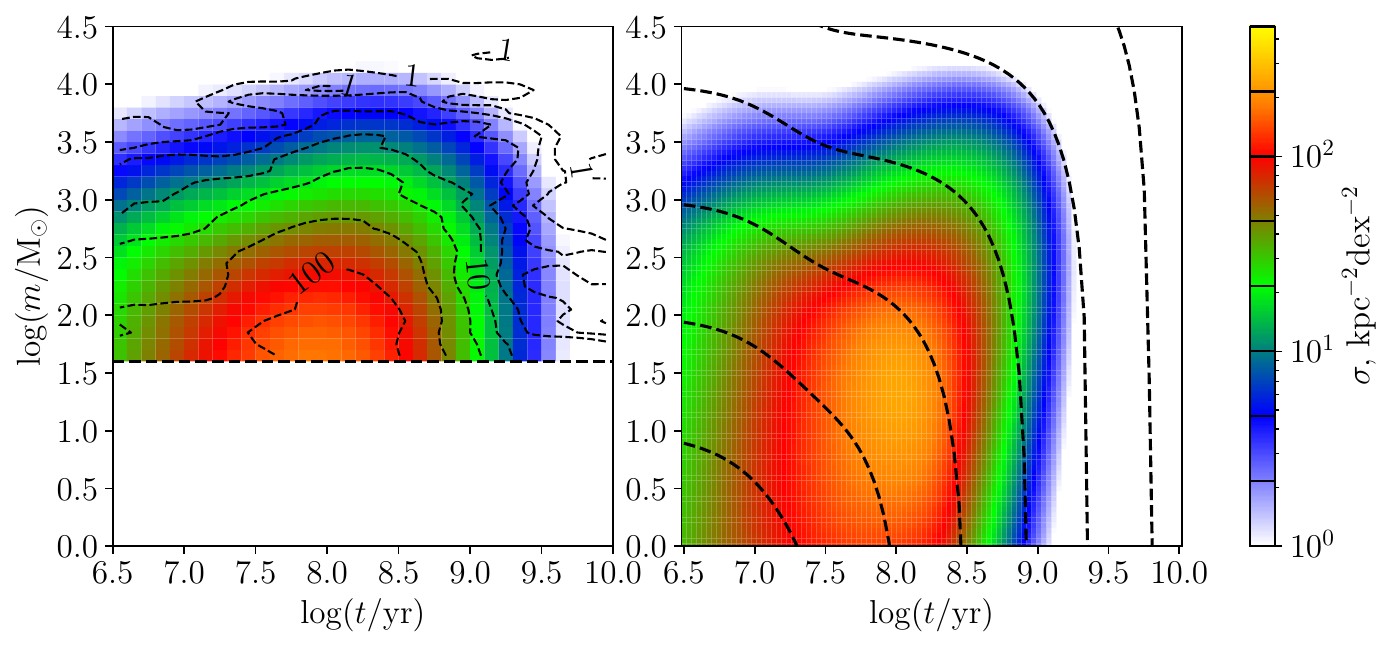}
    \caption{{Modelled CAMF for the fit to \citet{2023A&A...673A.114H, hunt&reffert24a} data.
    In the right panel, the model CAMF with no smoothing is shown.} Dashed lines are mass loss tracks corresponding to initial masses of $1\msun$, $10\msun$ , $10^2\msun$, $10^3\msun$, $10^4\msun$, $10^5\msun$, and $10^6\msun$. In the left panel, the contours of the smoothed observed CAMF are plotted over the error-smoothed model CAMF. Contour levels of the smoothed data are marked with solid lines on the colour scale bar.}
    \label{fig:hr24_model_camf}
\end{figure}

Indeed the best model, with its CAMF shown in Fig.~\ref{fig:hr24_model_camf}, does not feature a break in the cluster lifetime-mass relation and has a high value of $a_1$ at 0.538, resulting in slow dissolution at the end of clusters' lifetimes and a high number of clusters with masses below $40\msun$.
The Schechter cutoff mass at $\log m_\mrm{S} / \msun = 4.743$ corresponds to lifetimes close to the age limit and seems mostly sufficient to reproduce it.

The CFR implies that a majority of stars form as part of long-lived clusters, though a large fraction of stars become unbound from their parent cluster early in their life. Less than half the newly formed stellar mass remains bound past 10~Myr, and 85\% of the initial mass are lost within 100~Myr.
The bound fraction after violent relaxation is comparatively high at 34\%.

The model contours in the left panel of Fig.~\ref{fig:hr24_model_camf} appear to match the observed CAMF fairly well.
Synthetic cluster samples drawn from the model contain on average $2447 \pm 41$ clusters for a total surface density of $\Sigma_{\mu > 1.6} = 257.22 \pm 0.50 \mrm{kpc}^{-2}$ above the mass limit, which appears consistent with observations within the stochastic variation. However, all 5000 synthetic samples we drew have lower values of the KLD than the observed sample, indicating the presence of systematic differences between observation and model that cannot be explained by stochastic noise.

One may attempt to improve this model through a better error model, improved prior constraints of the model functions, a different fit statistic, or some combination thereof. 
We refrain at this point from a more in-depth analysis of this sample on the basis of its provisional nature and our primary aim of demonstrating the use of our model and methodology.

We find that the CAMFs of the MWSC and \citet{2023A&A...673A.114H, hunt&reffert24a} catalogue are qualitatively similar, in particular with respect to the mass structure for $\mu > 1.6$, the presence of a weakly mass-dependant age limit and the enhanced cluster mass loss at early ages. However, there are strong systematic differences in the age structure that cannot be explained by differences in age determination alone. As clusters grow fainter with age, they become more difficult to detect. Therefore it seems likely that there exist some unaccounted-for age-dependent cluster detection effects in at least the \citet{2023A&A...673A.114H, hunt&reffert24a} sample, which uses an approximate completeness distance based only on cluster mass. According to \citet{2023A&A...673A.114H} (see Fig.~15 therein), the fraction of MWSC clusters also found by their search decreases with age for $\tau > 8$. Further investigation of this catalogue's completeness limits will be important for future work making use of it. 

\section{Summary and conclusions}\label{sec:conc}

In this paper we investigated the parameter space of a theoretical model for the cluster age-mass function (CAMF), which was introduced in \citet{mwscmass}, where a base model was presented. The model was constructed with a constant cluster formation rate (CFR), a two-power law cluster initial mass function (CIMF) with an exponential Schechter cutoff, and a cluster bound mass function $m(M,t)$ describing the cluster mass loss as function of initial mass and age. The cluster mass evolution is adapted to the models of \citet{shukirea17,shukirea18} based on a centrally concentrated star formation efficiency. 

For finding the best model parameter set to reproduce the observed CAMF, we compared three different statistics, $\chi^2$, KLD, and maximum likelihood. All three resulting models are able to reproduce the bulk of the CAMF reasonably. However, only the KLD statistics returned a model which also fits the low number density regimes of the CAMF successfully, in particular in the {high-age and high-mass regimes}.

A posterior analysis by random sampling of the models was used to determine whether a configuration similar to the observed cluster sample is a likely configuration of such a model sample. For this, the different statistics as well as the total cluster number were used to quantify deviations between models and their samples, and assess whether the observed cluster sample can be differentiated from a model sample based on the observed values of the different statistics.
The base model, published in \citet{mwscmass}, can be differentiated from observations by $\chi^2$, KLD, and total cluster number, but not by the maximum likelihood method.
The models returned from $\chi^2$ and maximum likelihood can be differentiated less easily, but also show deviations in at least two of the statistics considered.
The model returned from the KLD in particular is best able to reproduce the observed cluster number and can only be differentiated from observations by the maximum likelihood method.

Using principal component analysis, we derived the ellipsoidal shape of the 11-dimensional posterior distribution in parameter space for the different MCMC runs. The semi-major axes of the ellipsoids (given by the square root of the {normalised covariance matrix} eigenvalues) vary by a factor of four, for example from ${\sqrt{\lambda_0}=0.341}$ to ${\sqrt{\lambda_{10}}=1.486}$ for the $D_\mrm{KL}$-fit (see also Fig.~\ref{fig:eigenvectors}). These eigenvalues give the variance of the parameter vector in the direction of the eigenvector relative to the individual parameter variances. An eigenvalue smaller than 1 corresponds to a combination of parameters that is more tightly constrained than the parameters are individually, while an eigenvalue larger than 1 corresponds to a combination that is constrained less so. The eigenvectors corresponding to both the smallest and largest eigenvalues are similar for the posteriors of the different runs, indicating that the different fit statistics realise the same constraints imposed by the observed CAMF. The model parameters are very well constrained in two dimensions fixing the upper age limit and the position of the maximum of the CAMF and badly constrained in one dimension corresponding to low-mass end of the distribution.

Finding tight constraints for individual model parameters and related physically interesting quantities is made difficult by these strong correlations between parameters. In several cases, these correlations make intuitive physical sense, and can be expected to generalise to other models. For instance, a model without a sharp break in the lifetime-mass relation requires a sufficiently small mass cut-off in order to reproduce the observed age limit.
A possible solution may be to restrict the input functions further by applying prior knowledge of the physical processes involved. The bound mass function and lifetime-mass relation are the best candidates for such a treatment, as the CIMF is already quite well-constrained and the shape of the CFR is very difficult to constrain both a priori and through our cluster sample.
While the CFR can be expected to generally correlate with the SFR, it is firstly not clear at all that there is a one-to-one relation between the two, and secondly there is as of yet no consensus as to the shape of the Galactic SFR \citep[see e.g.][]{aumebin09,2017MNRAS.470.1360B,sysjus21}.
Preliminary tests of models with an additional free parameter for an exponentially varying CFR have demonstrated that the observed cluster sample is compatible with both exponentially increasing and decreasing CFRs, but the results are not significant. The reason for this is a strong degeneracy with the cluster mass loss and lifetime parameters, which can compensate for moderate changes of the CFR shape.

The base model was constructed to reproduce the mass functions in four age bins \citep{mwscmass}. The focus was on the younger ages $\log t / \mrm{yr}<8.3$ and the maximum of the number surface density. The best model resulting from the $D_\mrm{KL}$-fit, which we adopted as our best-fitting model, provides a similarly good representation of these features of the CAMF and a significantly improved model for the old and the high-mass regions.

The observed CAMF shows a clear signature of a fast cluster mass loss in the violent relaxation phase as predicted by the N-body models of \citet{shukirea17,shukirea18,shukirgaliyev2019,shukirea21} with low global star formation efficiency. 
This mass loss depends chiefly on the details of cluster formation, in particular the star formation efficiency and the density profile of the embedded cluster.
However, poor number statistics for the youngest clusters and high tidal mass uncertainties limit how well the parameters of violent relaxation can be determined in our model.
In the best model, the bound fraction after violent relaxation is {28\%}, which corresponds to a global star formation efficiency around {18\%}.

The sharp mass-independent cutoff of the CAMF at high ages of $\sim 5\,$Gyr cannot be reproduced by the Schechter mass cutoff on its own, and requires a very weak dependence of the cluster lifetime on the initial cluster mass. 
{Comparisons with mass-lifetime parametrisations by \cite{ernstea15} suggest a possible explanation for this variation in the mass-lifetime relation: clusters with low initial masses may be born predominantly Roche-volume underfilling, while clusters with high initial masses may be  initially overfilling.

This lifetime boundary is connected to the strongest constraint in the 11-dimensional parameter space corresponding to the smallest eigenvalue $\lambda_0$.
It} may be relaxed if one allows for a CFR that decreases with age, in particular more rapidly than $1/t$ for $t > 1\,$Gyr. More investigation into relevant mechanisms that may limit cluster lifetimes, {a possible relation between initial filling factor and initial mass,} as well as the Milky Way's cluster forming history are needed to decide which scenario is more likely.

The CFR of the best model converts to a cluster formation rate in mass density of ${ 0.088 \msun \mrm{pc^{-2} Gyr^{-1}}}$, significantly smaller than in the base model due to generally slower cluster mass loss. This corresponds to {4$-$6.5\% contribution} {from observed clusters that survive gas expulsion} to the star formation rate of the disc.
{Depending on the completeness fraction of our sample, the total contribution of long-lived clusters to star formation is expected to be higher.  As discussed in Sect.~\ref{sec:gaia_data}, the cluster surface density constructed using \citet{2023A&A...673A.114H, hunt&reffert24a} is about three times higher than that of the MWSC. 
However, the fraction of stars bound in clusters is expected to drop rapidly with age due to the effects of violent relaxation and the dissolution of low-mass clusters, which contributes to the uncertainty of the CFR.
In our best model, half the initial stellar mass formed within bound clusters is lost within 30~Myr, and less than 20\% of the mass remains bound beyond the first 100~Myr.}

Overall, we find that the observed CAMF can be matched well by a relatively simple analytical model. 
{We demonstrate the robustness of our approach by further using the KLD for fitting and posterior analysis in conjunction with the Gaia-based catalogue of \citet{2023A&A...673A.114H, hunt&reffert24a}. We constructed the CAMF and notably found also the signature of enhanced mass loss during the violent relaxation phase. With minimal modification of our model, we were able to obtain a model describing also this cluster sample fairly well despite the fundamental differences in the underlying data.
We conclude that modelling the CAMF represents a general tool that can be used to extract information on the evolution of a cluster system from its observed age-mass structure.}

However, there are two main challenges in inferring model parameters and selecting best-fit models we identified over the course of this work. Both challenges are expected to not be specific to the model functions {and the cluster data} we used, but intrinsic to the problem of modelling the evolution of a cluster system. {We identified these first in the context of the MWSC dataset and found them to also be relevant when using the \citet{2023A&A...673A.114H, hunt&reffert24a} catalogue.} 
The first challenge lies in the strongly correlated nature of model parameter space, which makes constraining individual parameters difficult. For instance, the rates of cluster formation and dissolution must be balanced such that the model reproduces the observed cluster content, but individually each of these rates are badly constrained. A possible remedy for this problem may lie in better and more sophisticated constraints on cluster mass evolution from prior physics considerations such as N-body simulations.
The second challenge lies in the measuring of observation-model correspondence, where uncertainties in the measurements of observed cluster properties, the completeness limits of the observed cluster sample and the varying cluster density across the age-mass plane must be taken into consideration. In particular, the resolution of the observed CAMF in the age-mass plane is limited by both sample size and measurement uncertainties, which can make approaches that work without explicit construction of the observed CAMF, such as our expected log-likelihood approach, more attractive.
We demonstrate the use of forward modelling to check the behaviour of a fit statistic for a given model, and discuss how a model fitting procedure can be limited by the fit statistic used.
More work is needed to identify efficient statistics for measuring the correspondence of a CAMF model to a cluster sample{, especially in view of the extensive, high-quality cluster catalogues based on Gaia DR3 \citep{gaiadr3_23} that are currently emerging \citep[e.g.][]{2023A&A...673A.114H}.}

\begin{acknowledgements}
{We thank the anonymous referee for their insightful comments, which helped to make our arguments clearer and to improve several aspects of our discussion.}

This study was supported by the Deutsche Forschungsgemeinschaft (DFG, German Research Foundation) -- Project-ID 138713538 -- SFB 881 (``The Milky Way System'', subproject A06).

We would like to acknowledge the previous work with our colleagues A.~E.~Piskunov{, N.~V.~Kharchenko} and D.~A.~Kovaleva, who participated in the discussions at the beginning of this work.

PB and MI thank the support from the special program of the Polish Academy of Sciences and the U.S. National Academy of Sciences under the Long-term program to support Ukrainian research teams grant No.~PAN.BFB.S.BWZ.329.022.2023.

MI acknowledges the support by the Science Committee of the Ministry of Science and Higher Education of the Republic of Kazakhstan (Grant No.~BR21881880). 

The work of PB was also supported by the grant No. ~BR24992759 ``Development of the concept for the first Kazakhstan orbital cislunar telescope - Phase I'', 
financed by the Ministry of Science and Higher Education of the Republic of Kazakhstan. 

We acknowledge the use of the Simbad database, the VizieR Catalogue Service and other services operated at the CDS, France.

\end{acknowledgements}

\bibliographystyle{aa}
\bibliography{clubib}

\appendix

\section{Rapid mass evolution in the observed CAMF using linear age binning}\label{app:vr}
We argue in Sect.~\ref{sec:data} that the constant high-mass edge for $\tau < 7.8$ in the observed CAMF is evidence for rapid mass evolution during the violent relaxation phase.
One way to more accessibly visualise this is to remove the effect of logarithmic age binning by normalising using linear bin sizes in age instead (as also used in Fig.~\ref{fig:camf_imass} to visualise the factorisation of the CAMF in initial-mass).

We show the observed CAMFs with and without uncertainties in age and mass using linear age scaling in Fig.~\ref{fig:camf_age_linear}.
Without the effect of logarithmic age binning, the high-mass contours visually follow the change from rapid mass loss during violent relaxation to more moderate masss loss during later evolution, similar to the mass loss tracks in Fig.~\ref{fig:camfs_bestfit}.

\begin{figure}
    \centering
    \includegraphics[width=1\hsize]{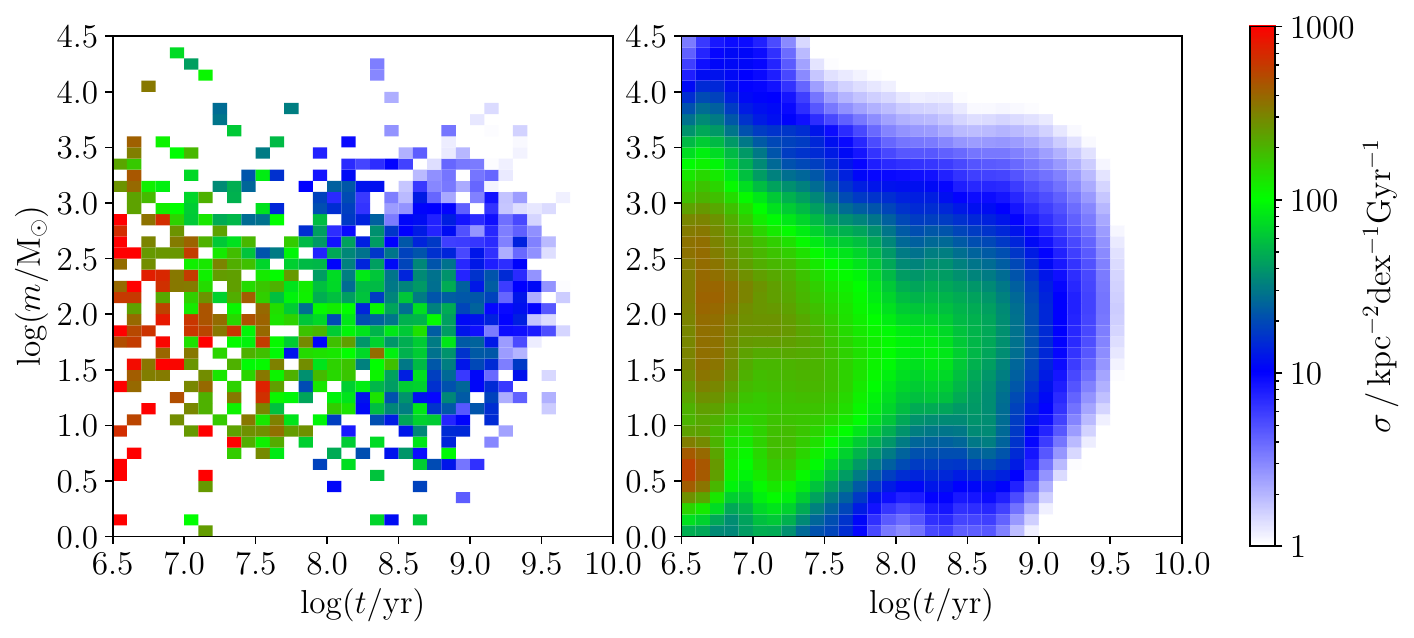}
    \caption{Observed CAMF constructed from MWSC data using logarithmic bins in mass and linear bins in age.
The left panel contains the CAMF constructed just from ages and masses, and the right panel contains the CAMF taking into account age-mass uncertainties.}
    \label{fig:camf_age_linear}
\end{figure}

\section{The modelled CAF}\label{sec:caf}

As is discussed in Sect.~\ref{sec:intro}, the CAF is typically of interest in the context of fitting cluster evolution models. While our 2D approach does not consider or fit the CAF explicitly, correspondence between observed and modelled CAF should nonetheless arise by construction.

For our model, the CAF $\eta(t)$, giving the cluster surface density per age interval, can be computed at some age $t$ by integrating the CAMF over all masses.
Recalling Eqs.~(\ref{eq:camf_differential}) and (\ref{eq:fsurv}), with $M_0(t)$ as inverse of $t_\mrm{Cl}(M)$, we find
\be
\eta (t) = \int\limits_0^\infty \sigma (m,t) \mrm{d}m =  \int\limits_{M_0(t)}^\infty \Psi (t) f(M) \mrm{d}M = \Psi (t) f_\mrm{surv}(t).
\ee
The CAF and CFR are thus simply related by the surviving fraction of clusters.

In Fig.~\ref{fig:cafs}, we plot $\eta(t)$ for the base model as well as the best-fitting models over the observed CAF.
As discussed in Sect.~\ref{sec:best_models}, the best model is the only one with reproduces the observed age limit. In terms of the CAF, this means that the other models produce high-age clusters in excess of observations.
Notably, the bulk of the CAF is very similar for all models, highlighting the additional information contained in the 2D structure of the CAMF, which is lost when only considering mass or age functions.

\begin{figure}
    \centering
    \includegraphics[width=0.9\hsize]{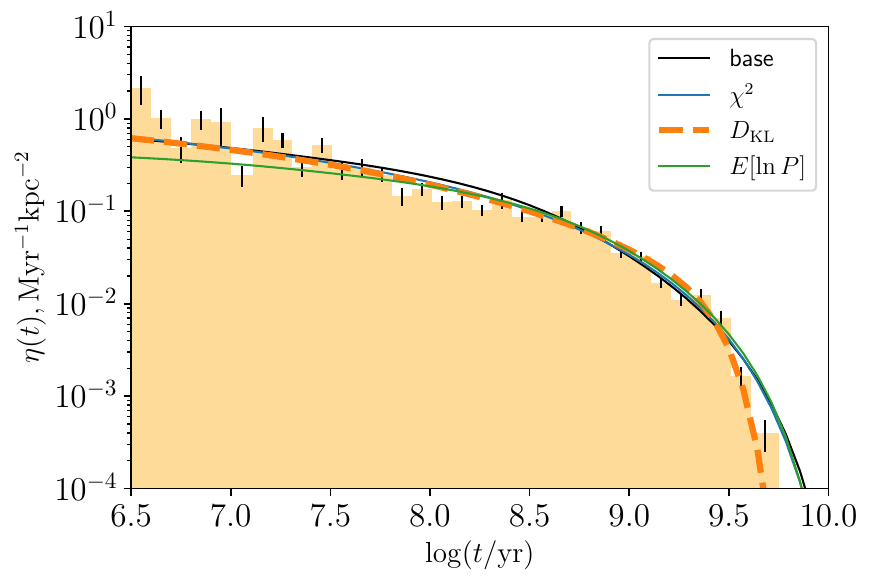}
    \caption{Observed and modelled CAFs. The histogram shows the observed CAF, error bars represent observational uncertainty from Poisson noise. Modelled CAFs are plotted for the base model, the $\chi^2$-fit, the $D_\mrm{KL}$-fit, and the $E[\ln P]$-fit model. The bold dashed line corresponds to the best model.}
    \label{fig:cafs}
\end{figure}

\end{document}